\documentclass[referee]{aa}

\usepackage{amsmath,amssymb}
\usepackage{natbib}
\bibpunct{(}{)}{;}{a}{}{,} 

\usepackage{graphicx}
\usepackage{bm}
\usepackage{color}
\newcommand{\revisednew}[1]{#1}
\newcommand{\revisedrenew}[1]{#1}

\begin{document}

\title{Clumpiness of Dark Matter and Positron Annihilation Signal}
\subtitle{Computing the odds of the Galactic Lottery}
\titlerunning{Positron annihilation in a clumpy halo}
\author{Julien Lavalle \inst{1} \and Jonathan Pochon \inst{2} 
  \and Pierre Salati \inst{3,4} \and Richard Taillet \inst{3,4}}
\authorrunning{Lavalle \textit{et al.}}

\institute{Centre de Physique des Particules CPPM, 
  CNRS-IN2P3 / Universit\'e de la M\'editerran\'ee, 
  13288 Marseille, France
  \and Laboratoire de Physique des Particules LAPP,
  74941 Annecy-le-Vieux, France
  \and Universit\'e de Savoie, 
  73011 Chamb\'ery, France 
  \and Laboratoire de Physique Th\'eorique LAPTH, 
  74941 Annecy-le-Vieux, France}
\offprints{R. Taillet}

\abstract{The small-scale distribution of dark matter in galactic
  halos is poorly known. 
  Several studies suggest that it could be very clumpy, which
  turns out to be of paramount importance when investigating
  the annihilation signal from exotic particles (e.g. 
  supersymmetric or Kaluza-Klein).}
{In this paper we focus on the annihilation signal
  in positrons. We estimate the associated uncertainty,
  due to the fact that we do not know exactly how
  the clumps are distributed in the Galactic halo.}
{To this aim, we perform a statistical study based on analytical
  computations, as well as numerical simulations. In particular, we 
  study the average and variance of the annihilation signal over many 
  Galactic halos having the same statistical properties.}
{We find that the so-called boost factor used by many authors should
  be handled with care, as i) it depends on energy and ii) it
  may be
  different for positrons, antiprotons and gamma rays, a fact which 
  has not received any attention before. As an
  illustration, we use our results to discuss the positron spectrum 
  measurements by the HEAT experiment.
}{}
\maketitle
\begin{flushright}
  preprint LAPTH-1142/06 and CPPM-P-2006-01
\end{flushright}

\section{Introduction}

Most observations of cosmological interest can be accounted for by
assuming that our Universe contains a large amount of non-baryonic
matter, usually referred to as \emph{dark matter}.
Indeed, the mean density of matter $\Omega_m$ can be consistently
estimated to be $\Omega_m \sim 0.23$ from many observations, whereas
the baryonic density $\Omega_b$ inferred from primordial
nucleosynthesis, from Cosmic Microwave Background (CMB) anisotropies
\citep{spergel2006},
Large Scale Structures
and by direct observations of luminous matter is an order of magnitude lower,
namely $\Omega_b h^2 \approx 0.0223$.

Among the several possible solutions to the dark matter problems, the
hypothesis that it could be made of a weakly interacting fundamental
particle of a new kind (hereafter wimp for weakly interactive massive
particle) has received considerable attention. This is partly due
to the fact that this hypothesis can be tested experimentally.
In particular, the detection of the annihilation
products of such exotic particles would be a great achievement, and
an important fraction of the astroparticle physics community is
involved in that quest.

However, assuming that wimps actually do exist  (see e.g. \citet{bertone}
for a nice review on dark matter), their nature is
basically unknown. Some constraints can be inferred from high
precision cosmological observations such as the CMB, but several particle
physics models provide candidates whose properties are consistent with
these observations. 
Extensions of the standard model of particle physics, such as
supersymmetry and Kaluza-Klein theories, naturally offer such candidates.
The lack of information about the nature of the wimp may translate into 
a very large uncertainty on its spatial distribution.
There are two main physical reasons for that.

\revisednew {First, the gravitational collapse of primordial density
  inhomogeneities that leads to the formation of cosmic structures is
  characterized by a small scale cut-off, due to several physical
  effects. To begin with, the particles enduring the collapse may
  interact elastically
  with other species or between themselves, which is responsible for
  diffusion. Then, after their interactions become negligible, they
  become free to move out of the collapsing region: this is
  free-streaming. A general discussion of these effects can be found
  in \citet{berezinsky03}. The resulting cut-off may strongly depend 
  on the nature and properties of the wimp (see e.g. \citet{boehm01,hofmann01}). 
  For instance, the
  recent study by \citet{profumo06} gives protohalo masses ranging from
  $3 \times 10^{-9}$ M$_\odot$ to $3 \times 10^{-1}$ M$_\odot$.}

\revisednew {Then, the structures evolve, merge and can be partially
  disrupted by tidal forces, so that the current cut-off in the 
  spectrum of clump masses corresponds to the smallest surviving
  clumps. The situation is still unclear, as
  numerical simulations by \citet{Diemand:2005vz} showed that
  clumps as small as $10^{-6}$ M$_\odot$ could survive disruption,
  while analytical work by \citet{berezinsky05} showed that structures
  smaller than $10^3$ M$_\odot$ were disrupted.
The possibility that tidal interaction with stars may play an
important role has also been hotly
debated \citep{zhao1, moore05, zhao2}.}

\revisednew{Different very competent experts of the domain provide
very different descriptions of the clumpiness of galactic halos, and
the aim of this paper is not to tackle with this issue. 
Instead, we consider the wide range of possibilities as the starting point of
our analysis.}
The quantitative amount of clumpiness is of paramount importance as it enhances the
annihilation rate of wimps and increases the detection prospects.

In most studies, clumpiness is taken into account by a general,
energy-independent multiplicative number called \emph{boost factor},
by which the signal computed from a smooth dark matter distribution
should be multiplied. 
This is not correct and we show in this paper that
the effects of clumpiness cannot be described by such a unique number.
Moreover, this is a stochastic problem, in the following sense: 
general hypotheses about the statistical properties of the
distribution of clumps in the Galactic halo can be made, 
but the exact position of every clump is unknown. In some cases,
the expected signal from a given type of wimp can be quite sensitive
to the precise position of the Earth relative to the nearest clumps.

The aim of this paper is to study the effect of the halo clumpiness on
the annihilation signal, focusing on the case of positrons.
Taking advantage of analytical computations and numerical simulations,
we investigate the statistical properties of the annihilation signal.
We show that, at variance with the assumptions of most studies, 
the clumpiness factor depends on energy and is not the same for
positrons as for gamma rays. We also show that even if the average
properties (averaging being meant over a large number of realizations
of our Galactic halo) of a clumpy halo may be well described by the usual
boost factor, the deviations from this average may be very large and
the ability to predict a signal from a model may be consequently deteriorated.

\revisednew{The importance of clumpiness in determining the
  dark matter annihilation signal in positrons has already been
  assessed by \citet{baltz_edsjo99}, and further studied 
  in \citet{hooper04a} and \citet{hooper04b}. 
  The possibility that the positron excess observed by HEAT could be 
  due to a single nearby clump had been raised. 
  The probability of such a situation was estimated to be low (about $10^{-4}$).
  More recently, this proposal resurfaces \citep{cumberbatch06}.
  As an illustration of the methods developed in this work, 
  we show that the odds for such an occurrence are even lower than
  Hooper et al's estimations.}


\section{The effective boost factor}
\label{sec:boost_effective}

For the sake of definiteness, 
we first consider the case of wimps annihilating into positrons
and electrons at a given energy -- the source spectrum of positrons
can be considered monoenergetic.
\revisednew{In Kaluza--Klein inspired models \citep{servant_tait},
dark matter species may substantially annihilate
into electron-positron pairs with a branching ratio as large as $\sim$ 20\%.}
The positron production rate $P_{e^+}$ counts the number of annihilations
taking place per unit volume at some point $\vec{x}$
\begin{equation}
  P_{e^+} \left( \vec{x} \right) = \delta \,
  \langle \sigma_{\rm ann}
  \left( \chi \, \chi \to e^+ \, e^{-} \right) \, v \rangle \,
  \left\{
    \frac{\rho \left( \vec{x} \right)}{m_{\chi}}
  \right\}^{2} 
  \label{annihilation_rate}
\end{equation}
where the $\delta$ term is equal to $1/2$ for Majorana particle,
taking into account the fact that these
particles are not discernible, whereas it is
equal to $1/4$ for Dirac particles, taking into account the fact that
the density of particles and antiparticles is $\rho/2$ and not $\rho$.
The contribution of the infinitesimal volume $d^{3} \vec{x}$ located
at point $\vec{x}$ to the flux at the Earth
\revisednew{-- in units of cm$^{-2}$ s$^{-1}$ sr$^{-1}$ GeV$^{-1}$ --}
of the resulting positrons with energy $E$ may be expressed as 
\begin{equation}
  d \phi = \mathcal{S} \,
  G_{e^+} \left( \vec{x}_{\odot} , E \leftarrow \vec{x} , E_{S} \right) \,
  \left\{
    \frac{\rho \left( \vec{x} \right)}{\rho_{0}}
  \right\}^2 \, d^3 \vec{x} \;\; ,
  \label{d_phi}
\end{equation}
where the quantity $\mathcal{S}$ depends on the mass density of reference
$\rho_{0}$ and on the specific features of the high energy physics model at
stake
\begin{equation}
  \mathcal{S} = \frac{\delta}{4 \pi} \, v_{e^+}(E) \,
  \langle \sigma_{\rm ann}
  \left( \chi \, \chi \to e^+ \, e^{-} \right) \, v \rangle \,
  \left\{
    \frac{\rho_{0}}{m_{\chi}}
  \right\}^2 \;\; .
\end{equation}
\revisednew{The velocity of the positron with energy $E$ at the Earth is denoted by $v_{e^+}$.}
The Green function $G_{e^+}$ is discussed in
section~\ref{subsec:positron_propagator}. It describes the probability
that a positron produced at point $\vec{x}$ with energy $E_{S}$
reaches the Earth with a degraded energy $E$.
As wimps are at rest
with respect to the Milky Way, the energy $E_{S}$ is equal to the parent
particle mass $m_{\chi}$.
At this stage, we would like to keep our discussion as general as
possible. Because our formalism should easily be extended to any charged
species -- to antiprotons or antideuterons for instance -- the positron
propagator will be denoted more simply as $G \left( \vec{x} , E \right)$.
The total positron flux at the Earth results from the integral over the
galactic DM mass distribution $\rho \left( \vec{x} \right)$
\begin{equation}
  \phi = \mathcal{S} \,
  \int_{\rm DM \, halo} \!\!
  G \left( \vec{x} , E \right) \,
  \left\{
    \frac{\rho \left( \vec{x} \right)}{\rho_{0}}
  \right\}^2 \, d^3 \vec{x} \;\; .
  \label{phi_1}
\end{equation}


Should the DM halo be 
smoothly distributed with mass density $\rho_{s}$,
the positron flux would be given by relation~(\ref{phi_1}) where the
wimp distribution is now described by $\rho_{s}$
\begin{equation}
  \phi_s = \mathcal{S} \,
  \int_{\rm DM \, halo}
  G \left( \vec{x} , E \right) \,
  \left\{
    \frac{\rho_{s} \left( \vec{x} \right)}{\rho_{0}}
  \right\}^2 \, d^3 \vec{x} \;\; .
\label{phi_s}
\end{equation}
\revisednew{
In the literature, the effects of clumpiness have been so far accounted
for by shifting upwards the flux $\phi_s$. The multiplicative factor is
called the boost. It acts as a constant of renormalization by which the flux
$\phi_s$ generated by a smooth DM halo should be multiplied in order to
take into account the enhancement of the wimp annihilation rate inside substructures.
That procedure has been widely used in the past but is shown to be wrong
in the present paper. In the following we discuss the method that must be followed
in order to compute correctly the signal $\phi$ at the Earth.}

\revisednew{
We assume that substructures -- whose density profile inside the i-th clump
is $\delta\rho_{i}(\vec{x})$ -- float inside a smoother background with mass
density ${\rho'}_{s}$ which is a priori different from $\rho_s$ introduced
above. The halo density $\rho$ can be written as
\begin{equation}
    \rho = {\rho'}_{s} \; + \; \sum_{i} \, \delta \rho_{i} \;\; ,
\label{sum_rho}
\end{equation}
and each clump has a mass
\begin{equation}
  M_{i} =
  \int_{\rm ith \, clump} d^{3} \vec{x} \;\;
  \delta \rho_{i} \left( \vec{x} \right) \;\; .
\end{equation}
Because wimp annihilation involves the square of the DM mass density,
the production of positrons inside the i-th protohalo is enhanced with
respect to the situation where that substructure would be diluted in
the surrounding medium. 
\revisedrenew{Should the latter be homogeneously spread
with a mass density $\rho_h$ (which will correspond to $f\rho_s$ below)},
the enhancement would be given by the boost factor $B_{i}$ which
we define as
\begin{equation}
  \int_{\rm ith \, clump} \!\!\!\!\! d^3 \vec{x} \;\;
  \delta \rho_{i}^2 \left( \vec{x} \right) = M_i \times 
  \rho_h \, B_i \;\; .
\end{equation}
That relation does not mean that the annihilation signal scales
linearly with the clump mass. The boost factor $B_{i}$ actually takes
into account the inner DM distribution so that various profiles for
$\delta \rho_{i}$ can lead to very different values for $B_{i}$. The
relevant quantity turns out to be the effective volume
$B_{i} \, M_{i} / \rho_h$. In the case of model (B) of
\citep{bertone2005} where the DM clumps have been accreted around
intermediate-mass black holes, the average value for that crucial
factor is $\sim 4 \times 10^{5}$ kpc$^{3}$ even if the spike radius
is only $\sim$ 1 pc.
Relation~(\ref{sum_rho}) translates into the positron flux at the Earth
\begin{equation}
    \phi = {\phi}'_s \, + \, \phi_{r} \;\; ,
\end{equation}
whose component
\begin{equation}
    \phi_{r} = \sum_{i} \; \varphi_{i}
\end{equation}
is produced by the constellation of DM protohalos that pervade the Milky
Way. The signal $\phi$ at the Earth is therefore enhanced by a factor of
$B \equiv \phi / \phi_s$ with respect to the situation where the DM
halo is completely smooth with mass density $\rho_{s}$.
Many clump distributions are possible and lead to different values
for the boost $B$. 
The distribution inside which we are embedded is of course unique. 
Unfortunately, we know little about it. 
In order to predict the set of plausible values for the boost
$B$, we are forced to consider the vast ensemble of all the possible DM
substructure configurations. 
Our lack of knowledge limits us to merely derive trends for the
boost. 
The analysis of how  $B$ is statistically
distributed is postponed to section~\ref{subsec:random_flux}. 
Instead we now focus on its average value $B_\text{eff}$ which suffices
when its variance is small.
To proceed further, a few simplifications are nevertheless helpful.}

\revisednew{
{\bf (i)} We will first assume that clumps are practically point-like.
\revisednew{This hypothesis is expected to be valid when the propagation distance is
  large compared to the size of the clump.}
As the volume of the galaxy filled by the DM substructures becomes
negligible, the halo density $\rho$ becomes
\begin{equation}
    \rho = {\rho'}_{s} \; + \; \sum_{i} \, M_{i} \;
    \delta^{3} \! \left( \vec{x} - \vec{x}_{i} \right) \;\; ,
\label{sum_rho_point_like}
\end{equation}
and the smooth component ${\phi}'_s$ of the flux is given by
relation~(\ref{phi_s}) where the mass density $\rho_{s}$ is now
replaced by ${\rho'}_{s}$. Moreover, the positron flux $\varphi_{i}$
which the clump located at position $\vec{x}_{i}$ yields, simplifies
into
\begin{equation}
  \varphi_{i} = \mathcal{S} \,
  \frac{B_{i} M_{i}}{\rho_{0}} \, G_{i} \;\; ,
\end{equation}
where $G_{i} \equiv G \left( \vec{x}_{i} , E \right)$.}

\revisednew{
{\bf (ii)} The boost factor $B_{i}$ at the source should vary from one
protohalo to another even if the mass $M_{i}$ is assumed constant.
To commence, the inner regions of the Milky Way have presumably collapsed
earlier than its outskirts, dragging with them substructures whose
concentrations are stronger than for the galactic periphery. We could
expect to have larger values of $B_{i}$ inside the solar circle.
However, clumps that move near the galactic center experience strong
tides that could significantly reshape them \citep{berezinsky03}. It
is not unconceivable that clumps partially evaporate like globular
clusters which exhibit characteristic tidal tails. If that effect is
dominant, the clump mass is reduced and probably the boost factor too
-- if the density profile of the substructure readjusts itself accordingly.
It is therefore quite difficult to predict how the clump boost factor
$B_i$ varies with position.
In order to simplify the discussion, we assume that all the clumps have
the same mass $M_{i} \equiv M_{c}$ and the same boost factor
$B_{i} \equiv B_{c}$. The first hypothesis is supported by numerical
simulations that indicate \citep{Diemand:2005vz}
that the mass function of substructures is a self-similar power law of slope
$dn(M) / d \log M \propto M^{-1}$ and is actually dominated by the lightest
clumps. The latter hypothesis is a priori more questionable
\citep{zhao1, moore05, zhao2, berezinsky05}. It is nevertheless a reasonable
choice insofar as the effects of a granular DM distribution on the flux of
positrons
will be shown to be mostly local. The actual value of $B_{i}$ should not
vary much in the solar neighbourhood and we can safely consider that it is
constant. The expression for the flux $\phi$ simplifies into
\begin{equation}
  \phi = {\phi}'_s + 
  \mathcal{S} \, \frac{B_{c} M_{c}}{\rho_{0}} \;
  \sum_{i} \; G_{i} \;\; .
  \label{phi_3}
\end{equation}}

\revisednew{
{\bf (iii)} A fraction $f$ of the total DM halo is in the form of
substructures embedded inside a smooth component with mass density
${\rho'}_{s}$.
In the intermediate-mass black hole scenario of \citet{bertone2005},
the fraction $f$ is so
small that ${\rho'}_{s} \simeq \rho_{s}$.
On the contrary, in the \citep{Diemand:2005vz} simulations, a value as large
as $f \sim 0.5$ is found with a preponderance of small-scale clumps which
should trace the smooth DM density as ascertained in \citet{berezinsky03}.
%
%
The mass density ${\rho'}_{s}$ could be quite different from $\rho_{s}$
but its contribution ${\phi}'_s$ to the overall signal $\phi$ is small.
We will assume for simplicity that
\begin{equation}
    {\rho'}_{s} = (1 - f) \, \rho_{s} \;\; ,
\end{equation}
where $f$ is constant all over the Milky Way. The corresponding flux ratio
${{\phi}'_s}/{\phi_s}$ -- which should not exceed 1 in any case -- is now
given by the factor $\left( 1 - f \right)^{2}$.}

\revisednew{
{\bf (iv)} A number $N_{H}$ of DM substructures pervade the Milky Way halo.
In this analysis, we will not consider the fluctuations of that number.
The probability that one of those lies at point $\vec{x}$ is controlled
by the distribution $p \left( \vec{x} \right)$. The number of clumps
which the volume $d^3 \vec{x}$ contains on average is
\begin{equation}
    \left\langle dn \right\rangle =
    N_{H} \, p \left( \vec{x} \right) \, d^3 \vec{x} \;\; .
\end{equation}
We infer an average flux at the Earth
\begin{equation}
  \langle \phi \rangle = \left( 1 - f \right)^{2} \, \phi_{s}  +
  \mathcal{S} \, \frac{B_{c} M_{c}}{\rho_{0}} \;
  \left\langle \sum_{i} \; G_{i} \right\rangle \;\; ,
  \label{phi_4}
\end{equation}
where the average sum over the Green functions $G_{i}$ is given by
the integral
\begin{equation}
  \left\langle \sum_i \; G_{i} \right\rangle =
  \int_{\rm DM \, halo} \!\!
  G \left( \vec{x} , E \right) \, \left\langle dn \right\rangle \;\; .
\end{equation}
For illustration purposes, we have chosen in our numerical examples
a particular clump distribution. Inspired by \citep{Diemand:2005vz},
we have assumed that protohalos trace the smooth distribution of
dark matter with
\begin{equation}
    p \left( \vec{x} \right) =
    {\displaystyle \frac{\rho_{s} \left( \vec{x} \right)}{M_{H}}} \;\; ,
\end{equation}
where $M_{H}$ is the mass of the DM Milky Way halo. We stress that our
analysis does not depend on that specific choice and is completely
general. Considering a different distribution
$p \left( \vec{x} \right)$ -- with no relation to the mass density
$\rho_{s}$ in particular -- would not qualitatively affect 
the main conclusions of our analysis.}

\revisednew{
Keeping this in mind, we can proceed with our illustrative choice for
$p \left( \vec{x} \right)$ and derive the effective boost
\begin{equation}
  B_\text{eff}(E) \equiv
  {\displaystyle \frac{\langle \phi \rangle}{\phi_{s}}} =
  \left( 1 - f \right)^{2} \; + \;
  f \, B_{c} \;
  {\displaystyle \frac{\mathcal{I}_{1}}{\mathcal{I}_{2}}} \;\; ,
  \label{B_eff_19}
\end{equation}
where the integral $\mathcal{I}_{n}$ is defined by
\begin{equation}
  \mathcal{I}_{n}(E) =
  \int_{\rm DM \, halo} \!\!
  G \left( \vec{x} , E \right) \,
  \left\{
    {\displaystyle \frac{\rho_{s} \left( \vec{x} \right)}{\rho_{0}}}
  \right\}^{n} \, d^{3} \vec{x} \;\; .
  \label{definition_I_cal_n}
\end{equation}}

\begin{figure*}[h!]
\centering
\includegraphics[width=0.45\linewidth]{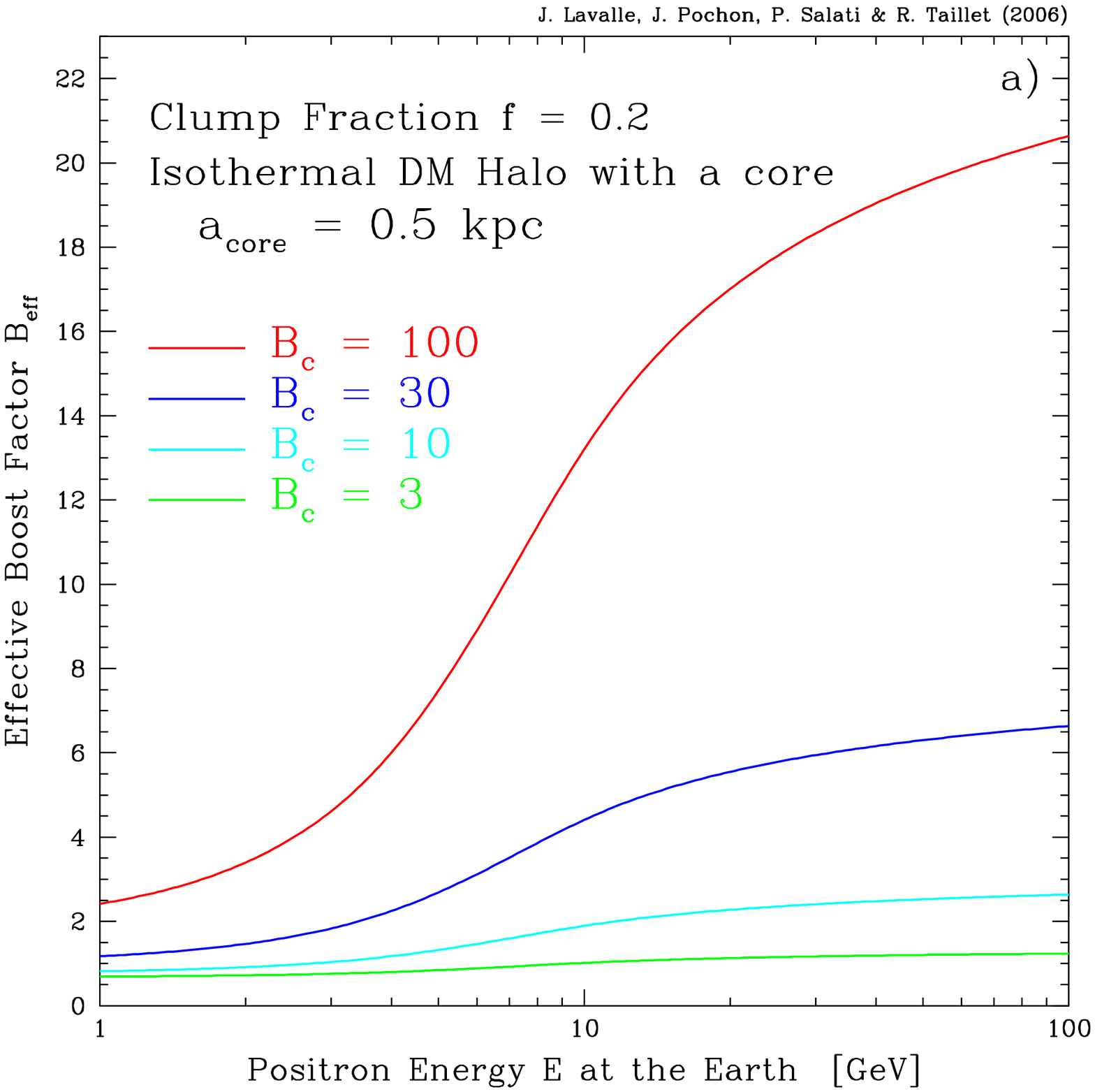}
\includegraphics[width=0.45\linewidth]{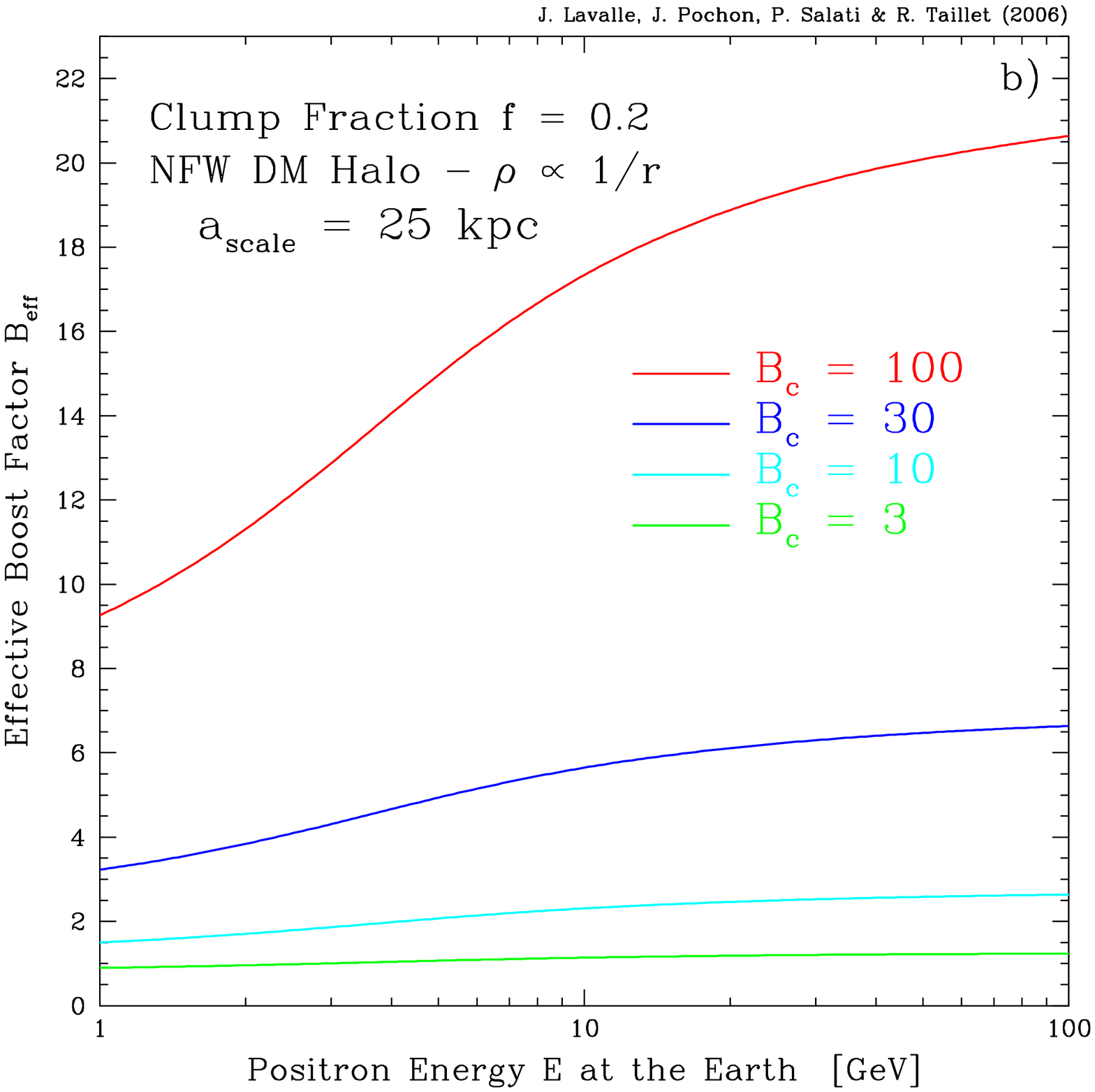}
\caption{
The effective boost factor $B_\text{eff}$ is featured as a function of the
positron energy $E$ in the case of a 100 GeV line. A fraction $f = 0.2$ of
the DM distribution is in the form of substructures whose individual boost
factor $B_{c}$ -- relative to the solar neighbourhood density -- has been
varied from 3 to 100. An isothermal halo -- panel a -- and a NFW
profile -- panel b -- are considered. They illustrate the influence
of the central profile index. The increase of $B_\text{eff}$ is
noticeable especially around $E \sim 10$ GeV.}
\label{fig:boost}
\end{figure*}

\revisednew{
Although the boost at the source $B_{c}$ is fixed, the boost of the
signal at the Earth $B_\text{eff}$ depends on both the nature and the energy
of the cosmic ray species through the Green function $G$ and the integrals
$\mathcal{I}_{1}$ and $\mathcal{I}_{2}$. We reach the conclusion that as
the flux $\phi_{s}$ is shifted upwards as a result of clumpiness, it also
experiences a spectral distorsion insofar as $B_\text{eff}$ is energy dependent.
This property has never been noticed before and is illustrated in the two
panels of Fig.~\ref{fig:boost} where the case of a 100 GeV positron
line is featured. A fraction $f = 0.2$ of the DM halo has collapsed in the
form of clumps whose boost factor $B_{c}$ is varied from 3 to 100.
In the \citep{Diemand:2005vz} numerical simulations, such a value for the
fraction would correspond to a minimum mass scale of $10^{4}$ M$_{\odot}$.
The latter lies typically at the lower tip of the range of protohalo masses
which we have used in our examples. As for the boost $B_{c}$, the values
quoted in the literature vary from a few \citep{berezinsky05} up to over two orders
of magnitude \citep{Diemand:2005vz}.}
%
%
The mass density of reference $\rho_{0}$ has been set equal to the solar
neighbourhood value of ${\rho_{s}(\odot)} = 0.3$ GeV cm$^{-3}$.
The increase of the effective boost factor with positron energy is clear
in both panels. Near the line -- in the region where $E$ tends to the
input energy $E_{S}$ -- the positron Green function $G$ probes
only a small region of the Milky Way halo around the solar system.
With our definition of $\rho_{0}$, the integral ratio
${\mathcal{I}_{1}}/{\mathcal{I}_{2}}$ boils down to unity and
$B_\text{eff}$ may be approximated by $\sim f \, B_{c}$.
If $E$ is now varied from its upper limit $E_{S}$ downwards, larger
portions of the halo come into play in the integrals ${\mathcal{I}_{1}}$
and ${\mathcal{I}_{2}}$ with the result of decreasing their ratio.
That effect is quite obvious in panel {\bf a)} where an isothermal
profile is assumed with core radius $a_{\rm core} = 0.5$ kpc. 
The DM density $\rho_{s}$
within 1 kpc from the galactic center is larger than in the case
of a NFW distribution and the relative increase of ${\mathcal{I}_{2}}$
-- where the square of $\rho_{s}$ is relevant and not merely
$\rho_{s}$ alone -- with respect to ${\mathcal{I}_{1}}$ is more
pronounced.
It is possible that the energy dependence of $B_\text{eff}$ which we have
discovered could strengthen the case of DM particles as a plausible
explanation of the still putative positron excess reported by HEAT.
In the example featured in Fig.~\ref{fig:boost}, the largest spectral
distortion is actually obtained for a positron energy $E \sim 10$ GeV.

\revisednew{Note that this distorsion effect should not be present in
  the case of gamma rays, whose propagation does not depend on
  energy. As regards antiprotons, the Green function already probes a
  significant portion of the DM halo, and we anticipate a mild
  dependence of the boost factor on the energy. 
} 
%

\section{The positron propagator}
\label{subsec:positron_propagator}

The departures of the positron flux from $\phi_{s}$ are expected to be large
when the positron energy $E$ is close to the production value $E_{S}$. In
this regime, the particles cannot have been produced far away. They mostly
originate from a region close to the solar system inside which the distribution
of clumps may significantly fluctuate. That is why we have focused our analysis
on cosmic ray positrons whose propagation throughout the galaxy is now briefly
sketched.

The master equation for positron propagation is the continuity relation
\begin{equation}
\partial_{\mu} J^{\mu} \, + \, \partial_{E} J^{E} = Q \;\; ,
\label{master_1}
\end{equation}
where $Q$ denotes the production rate of positrons per unit of volume and
energy. The space-time vector current is defined as
\begin{equation}
J^{\mu} =
{\displaystyle \frac{dn}{dE}} \;
\left\langle \dot{x}^{\mu} \right\rangle \;\; .
\end{equation}
The time-component
$J^{0} = {dn}/{dE} \equiv \psi \left( \vec{x} , E \right)$ denotes the
number density of particles per unit of volume and energy. The space current
accounts for the scattering of cosmic rays upon the inhomogeneities of
the galactic magnetic fields which is described as a diffusion process with
\begin{equation}
\vec{J} = - \, K \left( \vec{x} , E \right) \; \vec{\nabla} \psi \;\; .
\end{equation}
The energy component $J^{E}$ depends on the energy loss rate $b$ through
\begin{equation}
J^{E} = \psi \;
\left\{ \left\langle \dot{E} \right\rangle \equiv - \, b(E) \right\} \;\; .
\end{equation}
Above a few GeV, positron energy losses are dominated by synchrotron
radiation in the galactic magnetic fields and by inverse Compton
scattering on stellar light and on CMB photons. The energy loss rate
$b$ depends on the positron energy $E$ through
\begin{equation}
b(E) =
{\displaystyle \frac{E^{2}}{E_{0} \tau_{E}}} \;\; .
\end{equation}
We have set the energy of reference $E_{0}$ to 1 GeV and the typical energy
loss time is $\tau_{E} = 10^{16}$ s.
The master equation~(\ref{master_1}) may be expanded into
\begin{equation}
{\displaystyle \frac{\partial \psi}{\partial t}} \; - \;
\vec{\nabla} \cdot
\left\{ K \left( \vec{x} , E \right) \, \vec{\nabla} \psi \right\} \; - \;
{\displaystyle \frac{\partial}{\partial E}}
\left\{ b(E) \, \psi \right\} = Q \left( \vec{x} , E \right) \;\; .
\label{master_2}
\end{equation}
In order to simplify the discussion, steady state is assumed and the space
diffusion coefficient $K$ is taken to be homogeneous with the energy dependence
\begin{equation}
K \left( E \right) = K_{0} \,
\left\{ {\displaystyle \frac{E}{E_{0}}} \right\}^{\alpha} \;\; .
\end{equation}
The diffusion coefficient at 1 GeV is
$K_{0} = 3 \times 10^{27}$ cm$^{2}$ s$^{-1}$ with a spectral index of
$\alpha = 0.6$. The master equation~(\ref{master_2}) simplifies into
\begin{equation}
K_{0} \, \epsilon^{\alpha} \, \Delta \psi \; + \;
{\displaystyle \frac{\partial}{\partial \epsilon}}
\left\{
{\displaystyle \frac{\epsilon^{2}}{\tau_{E}}} \, \psi
\right\} \; + \; Q = 0 \;\; ,
\label{master_3}
\end{equation}
where $\epsilon$ denotes the ratio ${E}/{E_{0}}$.

Equation~(\ref{master_3}) may be solved with the \citet{baltz_edsjo99}
clever trick which consists in translating the energy $E$ into the pseudo-time
\begin{equation}
  \tilde{t}(E) = \tau_{E} \;
  \left\{
    v(E) \, = \, {\displaystyle \frac{\epsilon^{\alpha - 1}}{1 - \alpha}}
  \right\} \;\; .
  \label{connection_E_pseudo_t}
\end{equation}
The energy losses which positrons experience boil down to a mere
evolution in this pseudo-time so that the propagation
equation~(\ref{master_3}) greatly simplifies into
\begin{equation}
{\displaystyle \frac{\partial \tilde{\psi}}{\partial \tilde{t}}} \; - \;
K_{0} \, \Delta \tilde{\psi} =
\tilde{Q} \left( \vec{x} , \tilde{t} \, \right) \;\; .
\label{master_4}
\end{equation}
The space and energy positron density is now
$\tilde{\psi} = \epsilon^{2} \, \psi$ whereas the positron production rate
has become
$\tilde{Q} = \epsilon^{2 - \alpha} \, Q$. Notice that both $\tilde{\psi}$
and $\tilde{Q}$ have the same dimensions as before because $\epsilon$ is dimensionless.
Without any space boundary condition, equation~(\ref{master_4}) may be
readily solved. If a drop is deposited at the origin of the coordinates at
pseudo-time $\tilde{t}_{S} = 0$
\begin{equation}
\tilde{Q} \left( \vec{x}_{S} , \tilde{t}_{S} \, \right) =
\delta^{3} \! \left( \vec{x}_{S} \right) \delta \! \left( \tilde{t}_{S} \right)
\;\; ,
\end{equation}
the subsequent diffusion in an infinite 3D space would result into the
density $\tilde{\psi}$ at position $\vec{x}$ and pseudo-time $\tilde{t}$
given by the well--known Green function
\begin{equation}
\tilde{\psi} \left( \vec{x} , \tilde{t} \, \right) \equiv
\tilde{G} \left( \vec{x} , \tilde{t} \leftarrow \vec{0} , 0 \right) =
\theta \! \left( \tilde{t} \, \right) \,
\left\{ 4 \, \pi \, K_{0} \, \tilde{t} \, \right\}^{-3/2} \,
\exp \left\{ - \,
{\displaystyle \frac{r^{2}}{4 \, K_{0} \, \tilde{t} \,}}
\right\} \;\; ,
\label{propagator_reduced_3D}
\end{equation}
where $r \equiv \left| \vec{x} \right|$.
The general solution of equation~(\ref{master_4}) may be expressed
with the Green function $\tilde{G}$ as
\begin{equation}
\tilde{\psi} \left( \vec{x} , \tilde{t} \, \right) =
{\displaystyle \int_{\tilde{t}_{S} = 0}^{\tilde{t}_{S} = \tilde{t}}}
d \tilde{t}_{S} \;
\int d^{3} \vec{x}_{S} \;
\tilde{G} \left( \vec{x} , \tilde{t} \leftarrow \vec{x}_{S} , \tilde{t}_{S} \right) \,
\tilde{Q} \left( \vec{x}_{S} , \tilde{t}_{S} \, \right) \;\; ,
\end{equation}
and translates into
\begin{equation}
\psi \left( \vec{x} , E \right) =
{\displaystyle \int_{E_{S} = E}^{E_{S} = + \infty}} dE_{S} \;
\int d^{3} \vec{x}_{S} \;
G_{e^+} \left( \vec{x} , E \leftarrow \vec{x}_{S} , E_{S} \right) \,
Q \left( \vec{x}_{S} , E_{S} \, \right) \;\; .
\end{equation}
The positron propagator may be obtained from $\tilde{G}$ through
\begin{equation}
G_{e^+} \left( \vec{x} , E \leftarrow \vec{x}_{S} , E_{S} \right) =
{\displaystyle \frac{\tau_{E}}{E_{0} \, \epsilon^{2}}} \;
\tilde{G} \left( \vec{x} , \tilde{t} \leftarrow \vec{x}_{S} , \tilde{t}_{S} \right)
\;\; ,
\end{equation}
where the connection between the energy $E$ and the pseudo-time
$\tilde{t}$ is given by relation~(\ref{connection_E_pseudo_t}).
In the case of monochromatic positrons, the production rate is
\begin{equation}
Q \left( \vec{x} , E \, \right) =
P_{e^+} \! \left( \vec{x} \right) \, \delta \! \left( E - E_{S} \right) \;\; ,
\end{equation}
and the positron space and energy density at the Earth may be expressed
as
\begin{equation}
\psi \left( \vec{x}_{\odot} , E \right) =
\theta \! \left( E_{S} - E \right) \;
\int d^{3} \vec{x}_{S} \;
G_{e^+} \left( \vec{x}_{\odot} , E \leftarrow \vec{x}_{S} , E_{S} \right) \,
P_{e^+} \left( \vec{x}_{S} \right) \;\; .
\end{equation}
Equation~(\ref{d_phi}) is based on this relation.

The diffusive halo inside which cosmic rays propagate before escaping
into the intergalactic medium is pictured as a flat cylinder with
radius $R_{\rm gal} = 20$ kpc and extends along the vertical direction
from $z = - L$ up to $z = + L$. We have assumed here a half-thickness
of $L = 3$ kpc.
Without any boundary condition, the propagator $\tilde{G}$ would be
given by the 3D relation~(\ref{propagator_reduced_3D}). However,
cosmic rays may escape outside the diffusive halo and $\tilde{G}$
should account for that leakage.
In spite of the boundary at $R_{\rm gal}$, we have assumed that
cosmic ray diffusion is not limited along the radial direction but takes
place inside an infinite horizontal slab with thickness $2 L$. We have
nevertheless disregarded sources located at a radial distance $R$ larger
than $R_{\rm gal}$. 
Indeed, because their energy is rapidly degraded as they propagate,
positrons are observed close to where they are produced. Our radial
treatment is justified because positrons do not originate from far
away \citep{maurin_taillet03}.
Even in the case of antiprotons for which the galactic propagation range
is significantly larger than for positrons, the effects of the radial
boundary down at the Earth are not significant insofar as cosmic ray species
tend to leak above and beneath the diffusive halo at $z = \pm L$ instead
of traveling a long distance along the galactic disk.
The infinite slab hypothesis allows the radial and vertical directions to
be disentangled in the reduced propagator $\tilde{G}$ which may now be
expressed as
\begin{equation}
\tilde{G}
\left( \vec{x} , \tilde{t} \leftarrow \vec{x}_{S} , \tilde{t}_{S} \right)
=
{\displaystyle
\frac{\theta \! \left( \tilde{\tau} \right)}{ 4 \, \pi \, K_{0} \, \tilde{\tau}}}
\; \exp \left\{ - \,
{\displaystyle \frac{R^{2}}{4 \, K_{0} \, \tilde{\tau} \,}}
\right\} \;
\tilde{V}
\left( z , \tilde{t} \leftarrow z_{S} , \tilde{t}_{S} \right) \;\; ,
\end{equation}
where $\tilde{\tau} =  \tilde{t} - \tilde{t}_{S}$. The radial distance
between the source $\vec{x}_{S}$ and the point $\vec{x}$ of observation
is defined as
\begin{equation}
R =
\left\{
\left( x - x_{S} \right)^{2} \, + \,
\left( y - y_{S} \right)^{2}
\right\}^{1/2} \;\; .
\end{equation}
Should propagation be free along the vertical direction, the propagator
$\tilde{V}$ would be given by the 1D solution $\mathcal{V}_{1D}$ to
the diffusion equation~(\ref{master_4})
\begin{equation}
\tilde{V}
\left( z , \tilde{t} \leftarrow z_{S} , \tilde{t}_{S} \right)
\equiv
\mathcal{V}_{1D}
\left( z , \tilde{t} \leftarrow z_{S} , \tilde{t}_{S} \right) =
{\displaystyle \frac
{\theta \! \left( \tilde{\tau} \right)}
{\sqrt{ 4 \, \pi \, K_{0} \, \tilde{\tau}}}} \;
\exp \left\{ - \, {\displaystyle
\frac{\left( z - z_{S} \right)^{2}}{4 \, K_{0} \, \tilde{\tau} \,}}
\right\} \;\; .
\label{propagator_reduced_1D}
\end{equation}
But the vertical boundary conditions definitely need to be implemented.
Wherever the source inside the slab, the positron density vanishes
at $z = \pm L$.
A first approach relies on the method of the so-called  electrical
images and has been implemented by \citet{baltz_edsjo99}. Any point-like
source inside the slab is associated to the infinite series of its
multiple images through the boundaries at $z = \pm L$ which act as
mirrors. The n-th image is located at
\begin{equation}
z_{n} =
2 \, L \, n \; + \; \left( -1 \right)^{n} \, z_{S} \;\; ,
\end{equation}
and has a positive or negative contribution depending on whether $n$
is an even or odd number. When the diffusion time $\tilde{\tau}$ is
small, the 1D solution~(\ref{propagator_reduced_1D}) is a quite good
approximation. The relevant parameter is actually
\begin{equation}
\zeta =
{\displaystyle \frac{L^{2}}{4 \, K_{0} \, \tilde{\tau} \,}} \;\; ,
\label{definition_zeta}
\end{equation}
and in the regime where it is much larger than 1, the propagation
is insensitive to the vertical boundaries. On the contrary, when
$\zeta$ is much smaller than 1, a large number of images need to
be taken into account in the sum
\begin{equation}
\tilde{V}
\left( z , \tilde{t} \leftarrow z_{S} , \tilde{t}_{S} \right) =
{\displaystyle \sum_{n \, = \, - \infty}^{+ \infty}} \;
\left( -1 \right)^{n} \,
\mathcal{V}_{1D}
\left( z , \tilde{t} \leftarrow z_{n} , \tilde{t}_{S} \right) \;\; ,
\label{V_image}
\end{equation}
and convergence may be a problem.
It is fortunate that a quite different approach is possible in that
case. The 1D diffusion equation~(\ref{master_4}) actually looks like
the Schr\"{o}dinger equation -- in imaginary time -- that accounts for
the behaviour of a particle inside an infinitely deep 1D potential well
that extends from $z = - L$ to $z = + L$. The eigenfunctions of the
associated Hamiltonian are both even
\begin{equation}
\varphi_{n}(z) = \sin
\left\{ k_{n} \left( L - \left| z \right| \right) \right\}
\end{equation}
and odd
\begin{equation}
\varphi'_{n}(z) = \sin
\left\{ k'_{n} \left( L - z \right) \right\}
\end{equation}
functions of the vertical coordinate $z$. The wave-vectors $k_{n}$
and $k'_{n}$ are respectively defined as
\begin{equation}
k_{n} = \left( n - \frac{1}{2} \right)
{\displaystyle \frac{\pi}{L}} \;\; {\rm (even)}
\;\;\;\; {\rm and} \;\;\;\;
k'_{n} = n \, {\displaystyle \frac{\pi}{L}} \;\; {\rm (odd)} \;\; .
\end{equation}
The vertical propagator may be expanded as the series
\begin{equation}
\tilde{V}
\left( z , \tilde{t} \leftarrow z_{S} , \tilde{t}_{S} \right) =
{\displaystyle \sum_{n \, = \, 1}^{+ \infty}} \;\;
{\displaystyle \frac{1}{L}} \;
\left\{
e^{\displaystyle - \, \lambda_{n} \tilde{\tau}} \,
\varphi_{n} \left( z_{S} \right) \, \varphi_{n}(z)
\; + \;
e^{\displaystyle - \, \lambda'_{n} \tilde{\tau}} \,
\varphi'_{n} \left( z_{S} \right) \, \varphi'_{n}(z)
\right\} \;\; ,
\label{V_quantum}
\end{equation}
where the time constants $\lambda_{n}$ and $\lambda'_{n}$ are
respectively equal to $K_{0} \, {k_{n}}^{2}$ and $K_{0} \, {k'_{n}}^{2}$.
In the regime where $\zeta$ is much smaller than 1 -- for very large
values of the diffusion time $\tilde{\tau}$ -- just a few eigenfunctions
need to be considered for the sum~(\ref{V_quantum}) to converge.


\section{An analytic approach of the cosmic ray flux fluctuations}
\label{sec:statistical_analysis}

\subsection{The random flux $\phi_{r}$ and its variance}
\label{subsec:random_flux}

The cosmic ray flux~(\ref{phi_3}) at the Earth contains the random
component
\begin{equation}
  \phi_{r} = \sum_i \; \varphi_{i} =
  \mathcal{S} \, {\displaystyle \frac{B_{c} M_{c}}{\rho_{0}}} \,
  \sum_i \; G_{i} \;\; ,
  \label{phi_r}
\end{equation}
which is produced by the constellation of DM clumps inside the Milky Way
halo. Before embarking into our discussion, a few remarks are in order.

{\bf (i)} The actual distribution of DM substructures is of course
unique and so is the cosmic ray flux which it generates at the Earth.
We will nevertheless consider it as one particular realization among an
essentially infinite number of different other possible realizations.
We furthermore assume that clumps are distributed at random and that the
set of all their possible distributions makes up the statistical ensemble
which we consider in this section. 
The aim of our analysis is
to investigate how strongly the flux $\phi_{r}$
may fluctuate as a result of the random nature of the wimp
clump distribution. We will derive the associated cosmic-ray
flux variance $\sigma_{r}$ defined as
\begin{equation}
\sigma_{r}^{2} =
\langle \phi_{r}^{2} \rangle \, - \, \langle \phi_{r} \rangle^{2} \;\; .
\end{equation}
The variance $\sigma_{r}$ turns out to be an essential tool.
Because the total flux $\phi$ and its random component $\phi_{r}$
are shifted with respect to each other by the constant quantity
\begin{equation}
  \phi - \phi_{r}= \left( 1 - f \right)^{2} \, \phi_{s}  \;\; ,
  \label{phi_5}
\end{equation}
the variance of the former is given by
\begin{equation}
\sigma_{\phi}^{2} =
\langle \phi^{2} \rangle \, - \, \langle \phi \rangle^{2} =
\langle \phi_{r}^{2} \rangle \, - \, \langle \phi_{r} \rangle^{2} =
\sigma_{r}^{2} \;\; .
\end{equation}
The effective boost itself $B_\text{eff}$ which has been discussed
in section~\ref{sec:boost_effective} is an average value around
which the true flux enhancement $B \equiv \phi / \phi_{s}$ fluctuates
with the variance
\begin{equation}
  \sigma_{B} =
  {\displaystyle \frac{\sigma_{\phi}}{\phi_{s}}} =
  {\displaystyle \frac{\sigma_{r}}{\phi_{s}}} \;\; .
  \label{variance_sigma_B}
\end{equation}
Therefore, the determination of $\sigma_{r}$ leads immediately to
the boost fluctuations $\sigma_{B}$.

{\bf (ii)} We will furthermore assume that clumps are distributed
independently of each other. The problem is then greatly simplified
because we just need to determine how a single clump is distributed
inside the galactic halo in order to derive the statistical properties
of an entire constellation of such substructures. In particular, the
average value $\left\langle \phi_{r} \right\rangle$ of the random
component of the cosmic ray flux is readily obtained from the average
flux $\left\langle \varphi \right\rangle$ produced by a single clump
through the relation
\begin{equation}
\left\langle \phi_{r} \right\rangle = N_{H}
\left\langle \varphi \right\rangle \;\; ,
\label{average}
\end{equation}
where $N_{H}$ denotes the total number of clumps to be considered.
The variance $\sigma_{r}$ -- which is the crucial quantity as
regards the flux fluctuations -- may also be expressed as
\begin{equation}
  \sigma_{r}^{2} = N_{H} \, \sigma^{2} = N_{H}
  \left\{
    \langle \varphi^{2} \rangle \, - \,
    \langle \varphi \rangle^{2}
  \right\} \;\; .
  \label{variance}
\end{equation}

{\bf (iii)} The set of the random distributions of one single clump
inside the domain $\mathcal{D}_{H}$ forms the statistical ensemble
$\mathcal{T}$ which we need to consider. An event from that ensemble
consists in a clump located at position $\vec{x}$ within the
elementary volume $d^{3} \vec{x}$. Its probability $dP$ will be assumed
to follow the smoothed DM mass distribution $\rho_{s}$ so that
\begin{equation}
dP = p \left( \vec{x} \right) \, d^{3} \vec{x} =
{\displaystyle \frac{\rho_{s} \left( \vec{x} \right)}{M_{H}}} \,
d^{3} \vec{x} \;\; .
\end{equation}
The domain $\mathcal{D}_{H}$ over which our statistical analysis is
performed is so large that the total number $N_{H}$ of clumps which
it contains is essentially infinite. That region $\mathcal{D}_{H}$
behaves therefore like a so-called thermostat in statistical mechanics.
It encompasses of course the diffusive halo and may even be much bigger.
It may be thought -- but not exclusively -- as the entire Milky Way DM
halo. Its actual size has no importance because it will disappear from
the final results in the limit where the ratio $1 / N_{H}$ is negligible.
The only requirement is that $N_{H}$ should be much larger than the typical
number $N_{S}$ of clumps that effectively contribute to the signal $\phi_{r}$
at the Earth. The domain $\mathcal{D}_{H}$ contains the total DM mass
$M_{H}$ -- a fraction $f$ of which consists in $N_{H}$ identical clumps
so that
\begin{equation}
  N_{H} \, M_{c} = f \, M_{H} \;\; .
  \label{constitutive}
\end{equation}

We are now ready to derive the probability distribution
$\mathcal{P}(\varphi)$ associated to the signal $\varphi$ which a single
clump generates. The statistical properties of the random variable
$\varphi \left\{ \mathcal{T} \right\}$ translate those of the statistical
ensemble $\mathcal{T}$ itself. More precisely, the probability function
$\mathcal{P}(\varphi)$ is related to the space distribution
$p \left( \vec{x} \right)$ through
\begin{equation}
\mathcal{P} \left( \varphi \right) \, d\varphi = dP =
{\displaystyle \int_{\mathcal{D}_{\varphi}}} \,
p \left( \vec{x} \right) \, d^{3} \vec{x} \;\; .
\label{definition_mathcal_P_Phi}
\end{equation}
The subdomain $\mathcal{D}_{\varphi}$ over which the space distribution
$p \left( \vec{x} \right)$ should be integrated in the previous
expression yields a flux at the Earth comprised between $\varphi$ and
$\varphi + d\varphi$ ($\mathcal{D}_{H}$ is thus the union of all 
$\mathcal{D}_{\varphi}$).
 In the case of positrons, the probability
distribution $\mathcal{P}(\varphi)$ will be shown in
section~\ref{subsec:probability_density} to concentrate around
a flux $\varphi$ equal to 0.
The average value -- over the statistical ensemble $\mathcal{T}$ --
of any function $\mathcal{F}$ that depends on the flux $\varphi$ may
be expressed as
\begin{equation}
  \langle \mathcal{F} \rangle =
  \int \, \mathcal{F} \! \left( \varphi \right) \,
  \mathcal{P} \left( \varphi \right) \, d\varphi =
  {\displaystyle \int_{\mathcal{D}_{H}}} \,
  \mathcal{F} \left\{ \varphi \left( \vec{x} \right) \right\} \,
  p \left( \vec{x} \right) \, d^{3} \vec{x} \;\; .
  \label{average_F}
\end{equation}
%
\begin{figure}[h!]
  \centerline
  {\includegraphics[width=\linewidth]{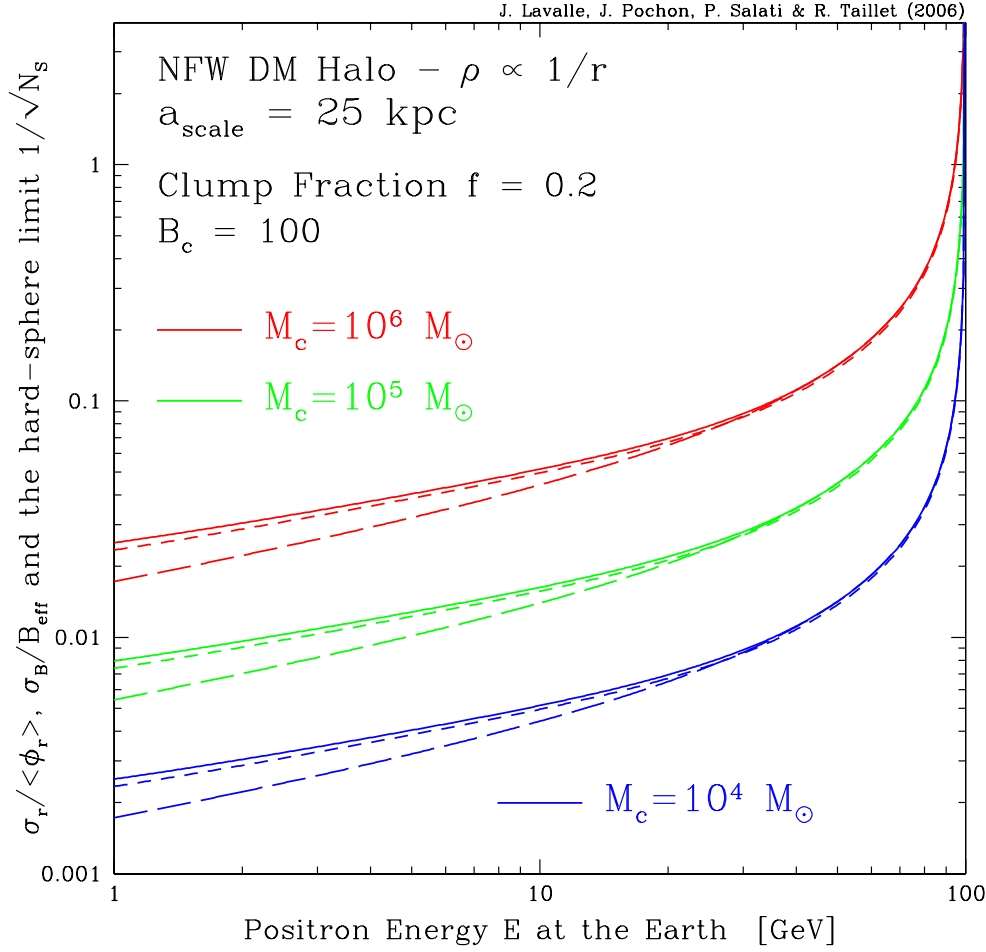}}
  \caption{
    The relative variance ${\sigma_{r}}/{\langle \phi_{r} \rangle}$ of the
    random component of the positron flux at the Earth -- solid lines -- and
    its hard-sphere approximation -- long-dashed curves -- are featured as
    a function of the positron energy $E$ for three different values of the
    clump mass $M_{c}$. The injected positron energy $E_{S}$ has been set
    equal to 100 GeV. A NFW profile with typical scale 25 kpc has been assumed.
    At fixed clump mass, the variance increases with $E$ and matches its
    hard-sphere approximation above $\sim$ 40 GeV.
    As the number of clumps is decreased, the curves are shifted upwards by a factor
    of $1/\sqrt{N_H} \propto \sqrt{M_c}$.
    The  relative variance
    ${\sigma_{B}}/{B_\text{eff}}$ of the boost factor is also displayed
    -- short-dashed curve. 
    In the limit where the clump boost factor $B_{c}$
    is large -- a value of 100 has been assumed here -- 
    ${\sigma_{B}}/{B_\text{eff}}$ and ${\sigma_{r}}/{\langle \phi_{r}
      \rangle}$
    are approximately equal.
}
  \label{fig:sigma_r_tot}
\end{figure}
%
In particular, the flux which a single clump yields on average
at the Earth is readily derived from the integral
\begin{equation}
\langle \varphi \rangle =
{\displaystyle \int_{\mathcal{D}_{H}}} \,
\varphi \left( \vec{x} \right) \,
p \left( \vec{x} \right) \, d^{3} \vec{x} =
\mathcal{S} \; {\displaystyle \frac{M_{c} B_{c}}{M_{H}}} \;
\mathcal{I}_{1} \;\; .
\end{equation}
where $\mathcal{I}_{n}$ has been defined in
relation~(\ref{definition_I_cal_n}). The average value of the random
flux $\phi_{r}$ implies $N_{H}$ clumps and expression~(\ref{average})
-- with the help of relation~(\ref{constitutive}) -- leads to
\begin{equation}
  {\displaystyle \frac{\langle \phi_{r} \rangle}{\phi_{s}}} =
  {\displaystyle \frac{\mathcal{S} f B_{c} \, \mathcal{I}_{1}}{\phi_{s}}}
  = f B_{c} \,
  {\displaystyle \frac{\mathcal{I}_{1}}{\mathcal{I}_{2}}} \;\; ,
\end{equation}
and to formula~(\ref{B_eff_19}).
%
\begin{figure}[h!]
  \centerline
  {\includegraphics[width=\linewidth]{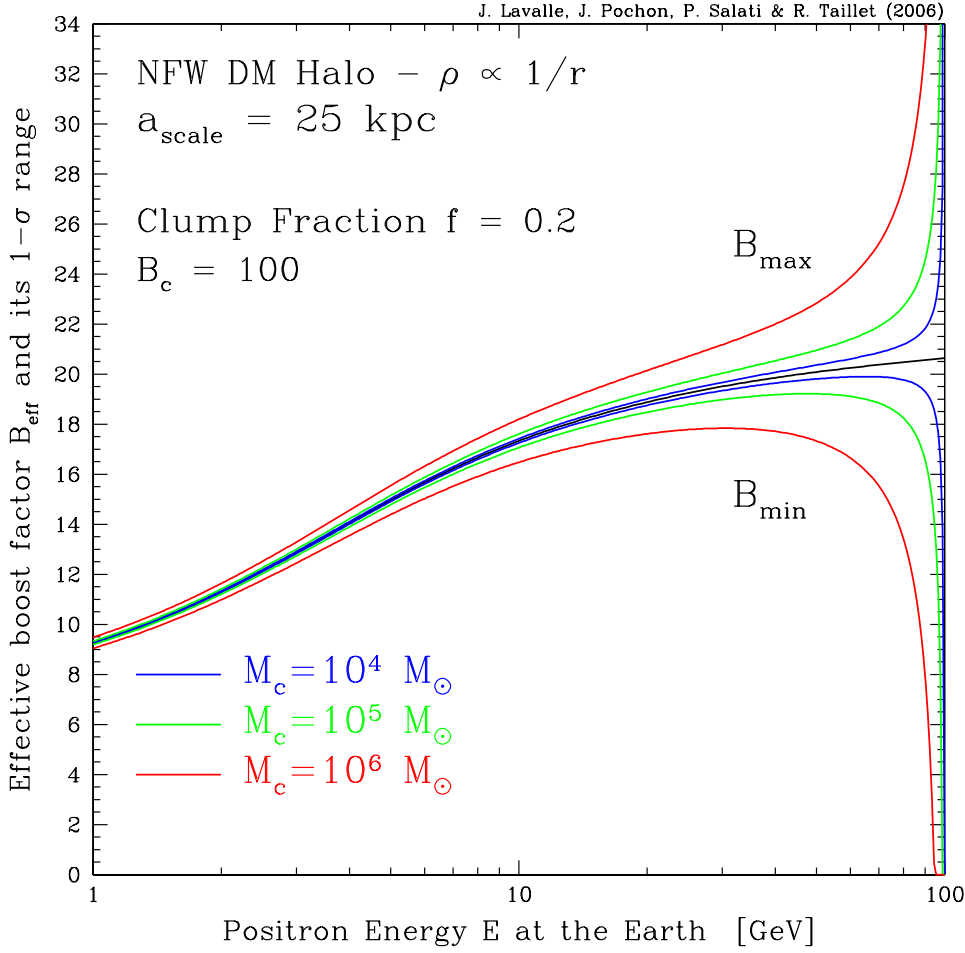}}
  \caption{
    The effective boost factor $B_\text{eff}$ -- black line -- is plotted as
    a function of the positron energy $E$ for an injected energy $E_{S} = 100$ GeV.
    The 1-$\sigma$ range of its fluctuations extends from
    $B_{\rm min} = B_\text{eff} - {\sigma_{B}}$ up to
    $B_\text{max} = B_\text{eff} + {\sigma_{B}}$.
    At fixed clump mass, that range opens up as $E$ approaches the injected
    energy $E_{S} = 100$ GeV.
    It also widens significantly at fixed positron energy $E$ when the
    number of clumps is decreased.}
  \label{fig:B_eff_and_its_range}
\end{figure}
%

Starting from the definition~(\ref{variance}), the variance $\sigma_{r}$
may be derived in the same spirit with
\begin{equation}
{\displaystyle \frac{\sigma_{r}^{2}}{\langle \phi_{r} \rangle^{2}}}
=
\frac{1}{N_{H}}
{\displaystyle \frac{\langle \varphi^{2} \rangle}{\langle \varphi \rangle^{2}}}
\, - \, \frac{1}{N_{H}} \;\; . 
\end{equation}
With the help of relation~(\ref{average_F}), the mean square of the
single clump flux may be expressed as
\begin{equation}
  \langle \varphi^{2} \rangle =
  \int_{\mathcal{D}_{H}} \,
  \varphi^{2} \! \left( \vec{x} \right) \,
  p \left( \vec{x} \right) \, d^{3} \vec{x} =
  \frac{\mathcal{S}^{2} M_{c}^{2} B_{c}^{2}}{\rho_{0} \, M_{H}} \,
  \mathcal{J}_{1} \;\; ,
\end{equation}
where the integral $\mathcal{J}_{n}$ is defined as
\begin{equation}
  \mathcal{J}_{n}(E) =
  \int_{\rm DM \, halo} \!\!
  G^{2} \left( \vec{x} , E \right) \,
  \left\{
    {\displaystyle \frac{\rho_{s} \left( \vec{x} \right)}{\rho_{0}}}
  \right\}^{n} \, d^{3} \vec{x} \;\; .
  \label{definition_J_cal_n}
\end{equation}
A straightforward algebra leads to the relative variance
\begin{equation}
  \frac{\sigma_{r}^{2}}{\langle \phi_{r} \rangle^{2}}
  = \frac{M_H}{\rho_0 N_H} \,
  \frac{\mathcal{J}_{1}}{\mathcal{I}_{1}^{2}}  - 
  \frac{1}{N_{H}}  \simeq \frac{M_{c}}{f \rho_{0}} \,
  \frac{\mathcal{J}_{1}}{\mathcal{I}_{1}^{2}} \;\; .
  \label{sigma_r_to_phi_r_F1}
\end{equation}
Because the domain $\mathcal{D}_{H}$ is so large -- remember that both
$\mathcal{D}_{H}$ and the Milky Way DM halo encompasses the diffusive
halo -- we can safely drop the ratio ${1}/{N_{H}}$ in the previous
expression.

The positron propagator of section~\ref{subsec:positron_propagator} has
been used in relation~(\ref{sigma_r_to_phi_r_F1}) in order to derive
the solid curves of Fig.~\ref{fig:sigma_r_tot}. 
At fixed $N_H$, the clump mass $M_{c}$ is determined by equation
\ref{constitutive} and the relative variance
${\sigma_{r}}/{\langle \phi_{r} \rangle}$ increases with the positron
energy $E$ at the Earth. This behaviour will be explained in
section~\ref{subsec:hard_sphere_discussion} with the hard-sphere
approximation. 
The ratio ${\sigma_{r}}/{\langle \phi_{r} \rangle}$ is proportional to
$1/\sqrt{N_H} \propto \sqrt{M_c}$, and weighted by an effective volume 
$\mathcal{J}_1/\mathcal{I}_1^2$.
The curves are therefore shifted upwards when the clump mass is
increased.
The relative variance $\sigma_B/{B_\text{eff}}$ of the flux
enhancement $B \equiv \phi / \phi_{s}$ is also presented in 
Fig.~\ref{fig:sigma_r_tot}. In the limit where the individual
clump boost factor $B_{c}$ is large -- we have selected here a value of
$B_{c} = 100$ -- the random component $\phi_{r}$ of the positron flux
dominates over its smooth counterpart
$\left( 1 - f \right)^{2} \, \phi_{s}$ so that
\begin{equation}
  {\displaystyle \frac{\sigma_{B}}{B_\text{eff}}} =
  {\displaystyle \frac{{\sigma_{r}}/{\phi_{s}}}
    {\left( 1 - f \right)^{2} \, + \, {\langle \phi_{r} \rangle}/{\phi_{s}}}}
  \, \simeq \,
  {\displaystyle \frac{\sigma_{r}}{\langle \phi_{r} \rangle}} \;\; .
  \label{relative_variances_equal}
\end{equation}
That is why the solid lines and short-dashed curves of Fig.~\ref{fig:sigma_r_tot}
are quite similar.
In Fig.~\ref{fig:B_eff_and_its_range}, the black central curve features
the effective boost factor $B_\text{eff}$ of a NFW halo and corresponds
to the case $B_{c} = 100$ of the panel b of Fig.~\ref{fig:boost} from
which it has been extracted. The 1-$\sigma$ range of its fluctuations
extends from
$B_{\rm min} = B_\text{eff} - {\sigma_{B}}$ up to
$B_\text{max} = B_\text{eff} + {\sigma_{B}}$. At fixed clump mass, that range
opens up as $E$ approaches the injected energy $E_{S} = 100$ GeV. The
fluctuations in the positron signal increase significantly just below the
positron line. 
The boost variance $\sigma_{B}$ is also proportional to $1/\sqrt{N_c}
\propto \sqrt{M_{c}}$. 
That is why the fluctuation band broadens
up as the clump mass is increased from $10^{4}$ up to $10^{6}$ M$_{\odot}$.

\subsection{The flux distribution $\mathcal{P} \! \left( \varphi \right)$ of a single clump}
\label{subsec:probability_density}

The positron flux at energy $E \leq E_{S}$ which a single clump located
at position $\vec{x}$ generates at the Earth implies the propagator discussed
in section~\ref{subsec:positron_propagator}
\begin{equation}
  \varphi \left( \vec{x} \right) =
  \mathcal{S} \, {\displaystyle \frac{B_{c} M_{c}}{\rho_{0}}} \;
  G_{e^+} \left( \vec{x}_{\odot} , E \leftarrow \vec{x} , E_{S} \right) \;\; .
\end{equation}
and may be expressed with the reduced Green function $\tilde{G}$ as
\begin{equation}
  \varphi \left( \vec{x} \right) =
  \mathcal{S} \, {\displaystyle \frac{B_{c} M_{c}}{\rho_{0}}} \,
  {\displaystyle \frac{\tau_{E}}{E_{0} \, \epsilon^{2}}} \;
  \tilde{G} \left( \vec{x}_{\odot} , \tilde{t} \leftarrow \vec{x} , \tilde{t}_{S} \right)
  \;\; .
\end{equation}
When the substructure is very close to the Earth, the flux $\varphi$ reaches
a maximal value $\varphi_\text{max}$ that depends both on the clump properties
through the \revisednew{effective volume $B_c M_c/\rho_0$} 
and on the specific features assumed for the DM particle through the factor
$\mathcal{S}$. Without any loss of generality, we can significantly
simplify the discussion by considering the ratio
\begin{equation}
  \Phi \left( \vec{x} \right) =
  {\displaystyle
    \frac{\varphi \left( \vec{x} \right)}{\varphi_\text{max}}} =
  {\displaystyle
    \frac{\tilde{G} \left( \vec{x} \right)}{\tilde{G}_\text{max}}} \;\; ,
\end{equation}
instead of the flux $\varphi$ itself. We therefore would like to derive
the density of probability
$\mathcal{P} \! \left( \Phi \right)$ associated to the reduced flux
$\Phi$ as it varies from 0 to 1.
%
\begin{figure}[h!]
  \centerline
  {\includegraphics[width=\linewidth]{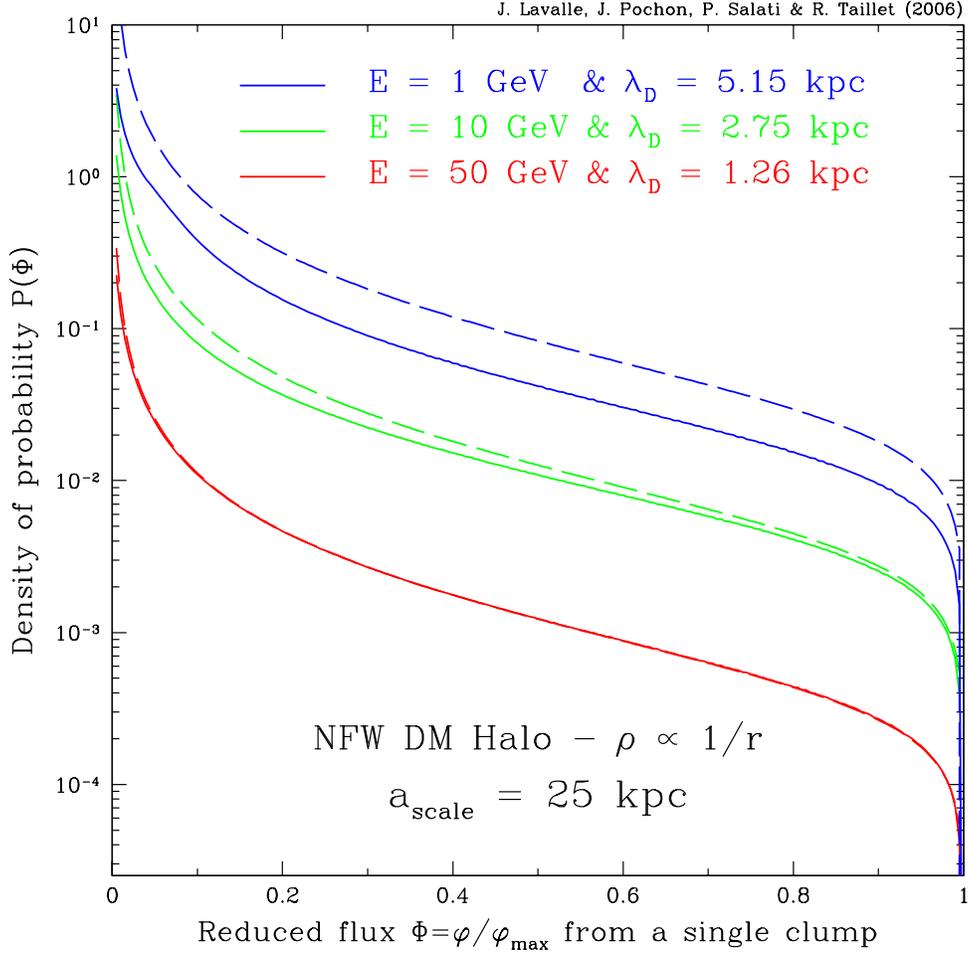}}
  \caption{
    The density of probability $\mathcal{P} \! \left( \Phi \right)$ is plotted
    as a function of the reduced flux $\Phi = \varphi / \varphi_\text{max}$
    which a single clump generates. A NFW halo has been assumed with a scale
    radius of 25 kpc. The domain $\mathcal{D}_{H}$ over which the
    probability is normalized to unity is the Milky Way DM halo up to a
    radius of 20 kpc. The injection energy is $E_{S} = 100$ GeV. The smaller
    the positron energy $E$ at the Earth, the larger the probability density
    for a non-vanishing flux. The fully numerical calculations -- solid curves
    -- are compared to the infinite 3D
    approximation~(\ref{analytic_mathcal_P_Phi}) that corresponds to the
    long-dashed lines.}
  \label{fig:probability_density_various_E}
\end{figure}
%
%
\begin{figure}[h!]
  \centerline
  {\includegraphics[width=\linewidth]{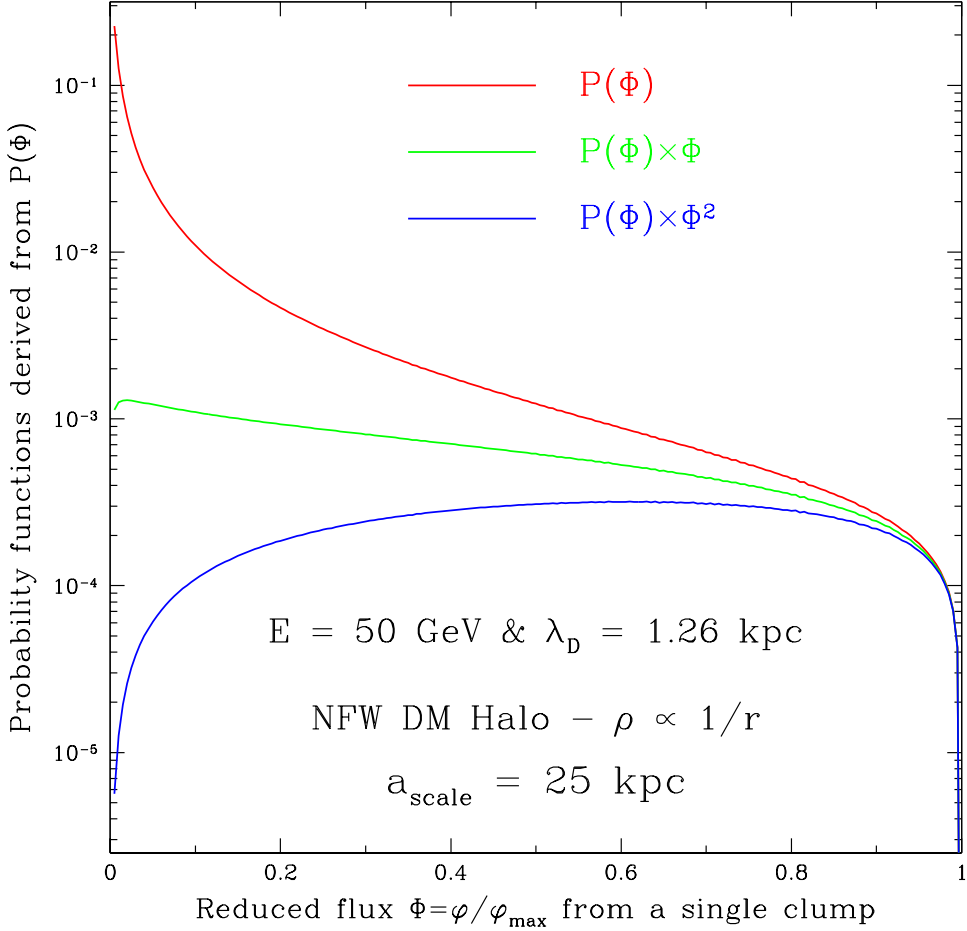}}
  \caption{
    The density of probability $\mathcal{P} \! \left( \Phi \right)$ as well
    as $\Phi \, \mathcal{P} \! \left( \Phi \right)$ and
    $\Phi^{2} \, \mathcal{P} \! \left( \Phi \right)$ are featured as a function
    of the reduced flux $\Phi = \varphi / \varphi_\text{max}$ for a positron
    energy at the Earth of 50 GeV.}
  \label{fig:probability_moments_E_50_GeV}
\end{figure}
%
%
\begin{figure}[h!]
  \centerline
  {\includegraphics[width=\linewidth]{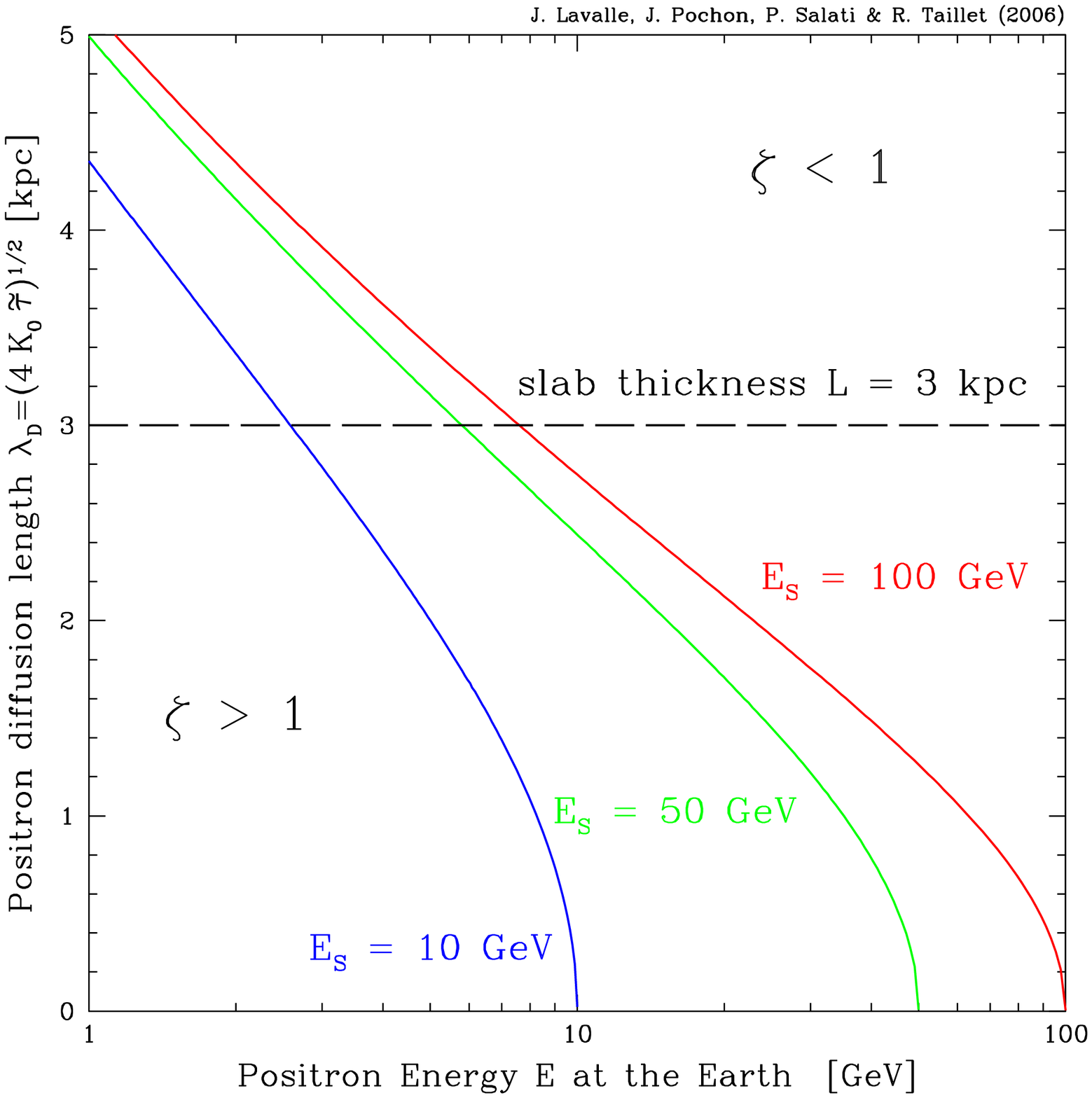}}
  \caption{
    The positron diffusion length $\lambda_{\rm D}$ decreases as the energy
    $E$ at the Earth approaches the energy $E_{S}$ of the line. The long-dashed
    horizontal line corresponds to a diffusion length $\lambda_{\rm D}$ equal to
    the thickness $L$ of the diffusion layers. Below that limit, positron
    propagation is not sensitive to the vertical boundaries and the infinite
    3D approximation is valid. This regime corresponds to large values of
    the parameter $\zeta$ -- see the definition~(\ref{definition_zeta}).}
  \label{fig:lambda_D_versus_E}
\end{figure}
%

In Fig.~\ref{fig:probability_density_various_E}, that distribution is
presented for three typical values of the positron energy $E$ at the
Earth. The energy $E_{S}$ of the positron line has been set equal to
100 GeV and a NFW DM halo has been assumed. The solid curves correspond
to the fully numerical calculation of $\mathcal{P} \! \left( \Phi \right)$
based on relation~(\ref{definition_mathcal_P_Phi}).
The domain $\mathcal{D}_{H}$ over which the probability is normalized
to unity is the Milky Way DM halo up to a radius of 20 kpc. That domain
encompasses the diffusive halo outside which the cosmic ray density vanishes. Most
of the probability is therefore contained in the low flux region and
the density $\mathcal{P} \! \left( \Phi \right)$ basically diverges at
$\Phi = 0$.
As the energy $E$ increases towards $E_{S}$, the region of the diffusive
halo that is probed by the positron propagator shrinks. That region
corresponds to large values of the positron flux $\Phi$. As its volume
decreases when $E$ approaches $E_{S}$, fewer clumps are involved in the
signal and the corresponding probability lessens. Notice in
Fig.~\ref{fig:probability_density_various_E} how the probability density
$\mathcal{P} \! \left( \Phi \right)$ drops when $E$ is increased from 1 GeV
up to 50 GeV.
The lower curve is reproduced in Fig.~\ref{fig:probability_moments_E_50_GeV}
together with the distributions $\Phi \, \mathcal{P} \! \left( \Phi \right)$
and $\Phi^{2} \, \mathcal{P} \! \left( \Phi \right)$ whose integrals
from $\Phi = 0$ up to $\Phi = 1$ are respectively related to
$\langle \varphi \rangle $ and $\langle \varphi^{2} \rangle$.

When the positron energy $E$ is close to the energy $E_{S}$, the
pseudo-time difference $\tilde{\tau} = \tilde{t} - \tilde{t}_{S}$ is so
small that the diffusion is no longer sensitive to the vertical boundaries
at $z = - L$ and $z = + L$. The Green function $\tilde{G}$ can be safely
approximated by the gaussian function (see \ref{propagator_reduced_3D})
\begin{equation}
  \tilde{G}
  \left( \vec{x}_{\odot} , \tilde{t} \leftarrow \vec{x} , \tilde{t}_{S} \right)
  =
  \left\{ 4 \, \pi \, K_{0} \, \tilde{\tau} \, \right\}^{-3/2} \,
  \exp \left\{ - \,
    {\displaystyle \frac{r^{2}}{4 \, K_{0} \, \tilde{\tau} \,}}
  \right\} \;\; .
  \label{infinite_3D_approximation}
\end{equation}
This regime corresponds to large values of the parameter $\zeta$ -- defined
in relation~(\ref{definition_zeta}) -- or alternatively to small values of
the positron diffusion length
$\lambda_{\rm D} \equiv \sqrt{4 \, K_{0} \, \tilde{\tau} \,}$. The latter
is featured in Fig.~\ref{fig:lambda_D_versus_E} as a function of $E$ for
three different values of the energy at source. In the case where
$E_{S} = 100$ GeV, the diffusion length $\lambda_{\rm D}$ exceeds the
thickness $L$ below an energy of $\sim 8$ GeV. Above that limit, positron
propagation is not affected by the vertical boundaries and the infinite 3D
approximation~(\ref{infinite_3D_approximation}) applies with a reduced
flux $\Phi$ that only depends on the distance $r$ of the clump to the Earth
\begin{equation}
  \Phi = \exp
  \left( - \, {\displaystyle {r^{2}}/{\lambda_{\rm D}^{2}}} \right) \;\; .
\end{equation}
An analytic density of probability may be derived in that case with
\begin{equation}
  \mathcal{P} \! \left( \Phi \right) = 2 \, \pi \, \lambda_{\rm D}^{3} \;
  {\displaystyle \frac{\rho_{s}(\odot)}{M_{H}}} \,
  {\displaystyle \frac{\sqrt{\displaystyle - \ln \Phi}}{\Phi}} \;\; .
\label{analytic_mathcal_P_Phi}
\end{equation}
That relation corresponds to the long-dashed curves of
Fig.~\ref{fig:probability_density_various_E} where a value of
$M_{H} = 1.357 \times 10^{11}$ M$_{\odot}$ has been found for
the mass contained in the inner 20 kpc of the Milky Way DM halo.
%
%
When the positron diffusion length $\lambda_{\rm D}$ is smaller than
the slab thickness $L$, relation~(\ref{analytic_mathcal_P_Phi}) is
an excellent approximation to the density of probability
$\mathcal{P} \! \left( \Phi \right)$. As an illustration, we find a value
of $\lambda_{\rm D} = 1.26$ kpc well below $L = 3$ kpc when the positron
energy $E$ is equal to 50 GeV. It is no surprise therefore if the solid
and long-dashed red lines of Fig.~\ref{fig:probability_density_various_E}
are so well superimposed on each other.
As $E$ decreases, the diffusion length $\lambda_{\rm D}$ becomes more
and more sizeable with respect to $L$ and the infinite 3D
propagator~(\ref{infinite_3D_approximation}) tends to overestimate
the region from which the signal originates as well as the corresponding
density of probability $\mathcal{P} \! \left( \Phi \right)$. Notice
how the long-dashed approximation lines are shifted upwards
with respect to the solid true numerical curves in
Fig.~\ref{fig:probability_density_various_E}. As $E$ decreases, the
approximation~(\ref{analytic_mathcal_P_Phi}) worsens and the
disagreement with the correct result becomes more pronounced.

\section{The hard-sphere approximation}
\label{subsec:hard_sphere_discussion}

In the limit where the infinite 3D approximation applies -- actually
for a large range of values of the positron energy $E$ at the Earth --
we can simplify further the propagator $G_{e^+}$ and substitute
the step function
\begin{equation}
  \tilde{G}
  \left( \vec{x}_{\odot} , \tilde{t} \leftarrow \vec{x} , \tilde{t}_{S} \right)
  =
  {\displaystyle
    \frac{\theta \! \left( r_{S} - r \right)}{V_{S}}}
  \label{approximation_G_hard_sphere}
\end{equation}
for the gaussian form~(\ref{infinite_3D_approximation}).
The distance between the clump and the
Earth is denoted by $r \equiv \left| \vec{x} - \vec{x}_{\odot} \right|$.
According to this hard-sphere approximation, the Green function $\tilde{G}$
reaches the constant value ${1}/{V_{S}}$ inside the sphere $\mathcal{D}_{S}$
of radius $r_{S}$ and volume $V_{S}$ -- whose center coincides with the Earth
-- and vanishes elsewhere. Both expressions~(\ref{infinite_3D_approximation})
and (\ref{approximation_G_hard_sphere}) are normalized to unity. The integral
over the full 3D space of the square of those Green functions should also be
the same. This condition translates into
\begin{equation}
  \frac{1}{V_{S}} = \int \, \tilde{G}^{2} \, d^{3} \vec{x} \;\; ,
\end{equation}
and leads to the volume
\begin{equation}
  V_{S} = \left( \sqrt{2 \pi} \, \lambda_{\rm D} \right)^{3} \;\; .
\end{equation}

In spite of its crudeness, the hard-sphere approximation turns out to be
quite powerful and is an excellent tool to understand the salient features
of the statistical properties of the clump distribution and of its flux.
The associated density of probability has little to do with the curves
of Fig.~\ref{fig:probability_density_various_E} or with
relation~(\ref{analytic_mathcal_P_Phi}). It is actually a bimodal
distribution with
\begin{equation}
  \mathcal{P} \! \left( \Phi \right) =
  p \, \delta \! \left( \Phi - 1 \right) \; + \;
  \left( 1 - p \right) \, \delta \! \left( \Phi \right) \;\; .
\end{equation}
The reduced flux $\Phi$ takes the value of 1 inside the sphere
$\mathcal{D}_{S}$ and is equal to 0 outside. The probability $p$ that
a clump lies inside the domain $\mathcal{D}_{S}$ -- from which it may
yield a signal at the Earth -- is just the ratio $M_{S} / M_{H}$ of the
DM mass $M_{S}$ confined in that sphere with respect to the DM mass $M_{H}$
contained in the entire domain $\mathcal{D}_{H}$. In the limit where
$\lambda_{\rm D} \propto r_{S}$ is small, the DM distribution is homogeneous
inside the sphere $\mathcal{D}_{S}$ -- with constant density
${\rho_{s}(\odot)}$ -- and the probability $p$ may be expressed as the ratio
\begin{equation}
p =
{\displaystyle \frac{M_{S}}{M_{H}}} =
{\displaystyle \frac{V_{S} \, {\rho_{s}(\odot)}}{M_{H}}} \;\; .
\end{equation}
%
%
For an injected energy $E_{S} = 100$ GeV and a positron energy at the Earth
$E = 50$ GeV, we find a probability $p \sim 2 \times 10^{-3}$ when the
statistical domain $\mathcal{D}_{H}$ is chosen to be the above-mentioned
NFW halo extending up to 20 kpc from the center of the Milky Way.

Because $p$ is vanishingly small and the number of clumps $N_{H}$ inside
the domain $\mathcal{D}_{H}$ exceedingly large, the limit of Poisson
statistics is reached. The probability to find $n$ clumps inside the sphere
$\mathcal{D}_{S}$ is therefore given by
\begin{equation}
P(n) =
{\displaystyle \frac{{N_{S}}^{n}}{n!}} \,
\exp \left( - N_{S} \right) \;\; ,
\label{poisson_1}
\end{equation}
where $N_{S} \equiv p \, N_{H}$ is the average number of clumps that
contribute to the signal
\begin{equation}
\langle n \rangle = N_{S} =
{\displaystyle \frac{V_{S} f {\rho_{s}(\odot)}}{M_{c}}} \;\; .
\end{equation}
Departures from the statistical law~(\ref{poisson_1}) in the case of
a realistic positron propagator will be discussed in
section~\ref{subsec:theorem_central_limit} when the number $N_{S}$
of the clumps involved in the flux at the Earth is large whereas the
opposite regime will be addressed in section~\ref{sec:limit_small_N_S}.
The Poisson distribution~(\ref{poisson_1}) is associated to the variance
\begin{equation}
\sigma_{n}^{2} =
\langle n^{2} \rangle \, - \, \langle n \rangle^{2} = N_{S} \;\; .
\end{equation}
In the hard-sphere approximation, the random part $\phi_{r}$ of the
positron flux at the Earth -- the contribution which the entire constellation
of substructures generates -- is proportional to the number $n$ of clumps
lying inside the sphere $\mathcal{D}_{S}$. We therefore anticipate that
the relative variance ${\sigma_{r}}/{\langle \phi_{r} \rangle}$ should be
equal to the relative variance ${\sigma_{n}}/{\langle n \rangle}$ of the
Poisson law~(\ref{poisson_1}).
As a matter of fact, in the limit where $\lambda_{\rm D}$ is small with
respect to $L$ -- and where the hard-sphere approximation becomes valid --
the integrals $\mathcal{J}_{1}$ and $\mathcal{I}_{1}$ simplify. If the
mass density of reference $\rho_{0}$ is set equal to its solar neighbourhood
value ${\rho_{s}(\odot)}$, the ratio
${\mathcal{J}_{1}}/{\mathcal{I}_{1}^{2}}$ boils down to ${1}/{V_{S}}$
so that the exact relation~(\ref{sigma_r_to_phi_r_F1}) simplifies into
\begin{equation}
{\displaystyle \frac{\sigma_{r}^{2}}{\langle \phi_{r} \rangle^{2}}}
=
{\displaystyle \frac{M_{c}}{f \rho_{0}}} \,
{\displaystyle \frac{\mathcal{J}_{1}}{\mathcal{I}_{1}^{2}}} =
{\displaystyle \frac{M_{c}}{V_{S} f {\rho_{s}(\odot)}}} =
{\displaystyle \frac{1}{N_{S}}} \;\; .
\label{sigma_r_to_phi_r_F2}
\end{equation}
We have therefore shown that in the hard-sphere regime, the variance
$\sigma_{r}$ of the random flux $\phi_{r}$ is indeed given by the
variance $\sigma_{n}$ that characterizes the Poisson
statistics~(\ref{poisson_1}). In Fig.~\ref{fig:sigma_r_tot},
the relative variance ${\sigma_{r}}/{\langle \phi_{r} \rangle}$ -- solid
curves -- and its hard-sphere approximation ${1}/{\sqrt{N_{S}}}$ --
long-dashed lines -- are presented together for comparison. Above
a positron energy at the Earth of 40 GeV, the correct calculation
and its hard-sphere limit differ by less than $\sim 5 \times 10^{-3}$.
%
%
The agreement is remarkable. The diffusion length does not exceed
$\sim 1.5$ kpc in that case and the hard-sphere approximation 
successfully describes the statistical properties of the random positron flux $\phi_{r}$.
The relative variance
${\sigma_{B}}/{B_\text{eff}}$ of the boost factor is also well reproduced
by the hard-sphere approximation ${1}/{\sqrt{N_{S}}}$ and both the short-dashed
and long-dashed curves are hardly distinguishable from each other at high
positron energy $E$.

\section{A Monte-Carlo approach of the cosmic ray flux fluctuations}
\label{subsec:MC_analysis}

\subsection{The large $N_{S}$ limit and the central limit theorem}
\label{subsec:theorem_central_limit}

%
\begin{figure}[th]
  \centerline{\hfill\includegraphics[width=0.5\linewidth]{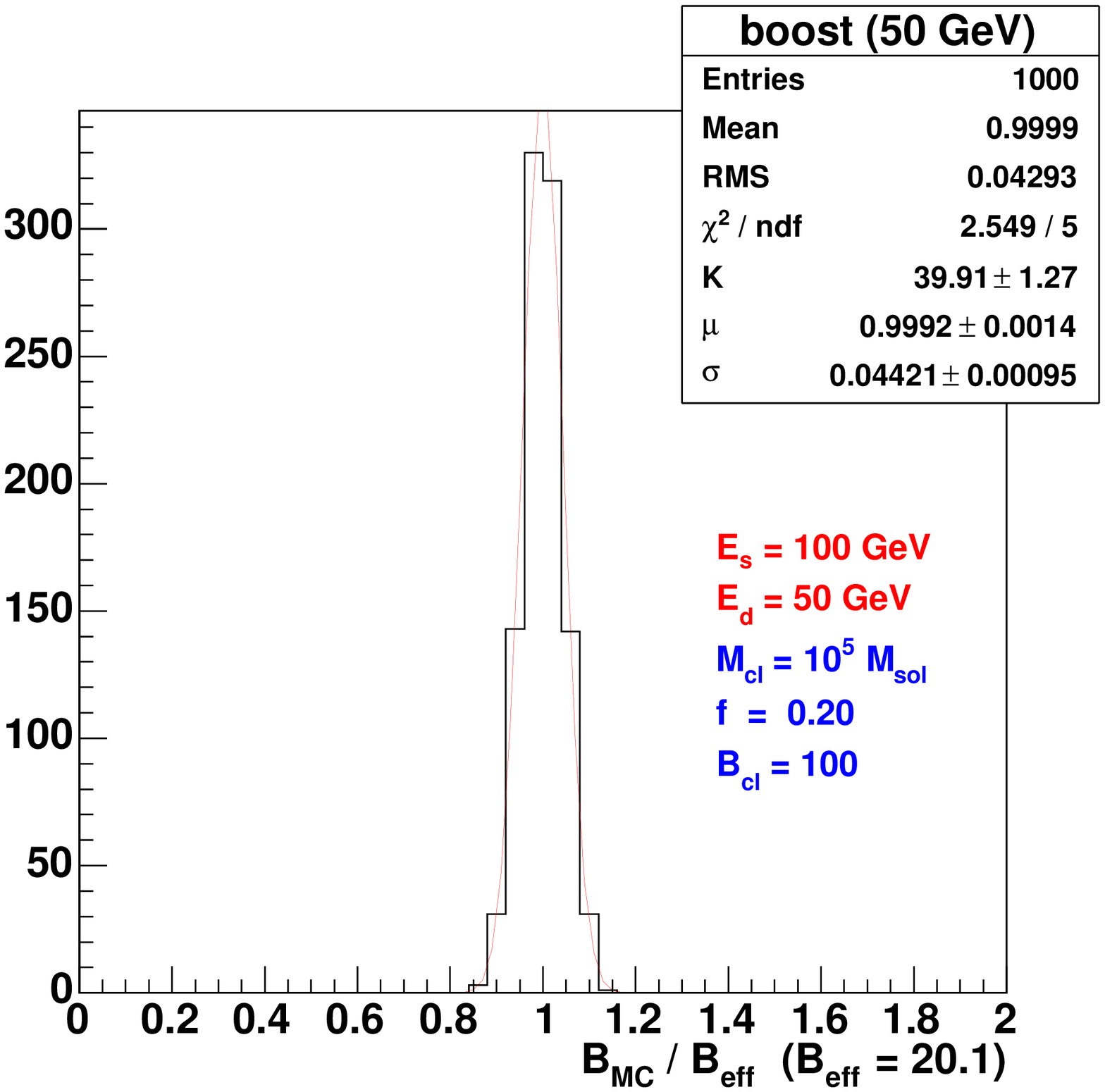}\hfill
    \hfill\includegraphics[width=0.5\linewidth]{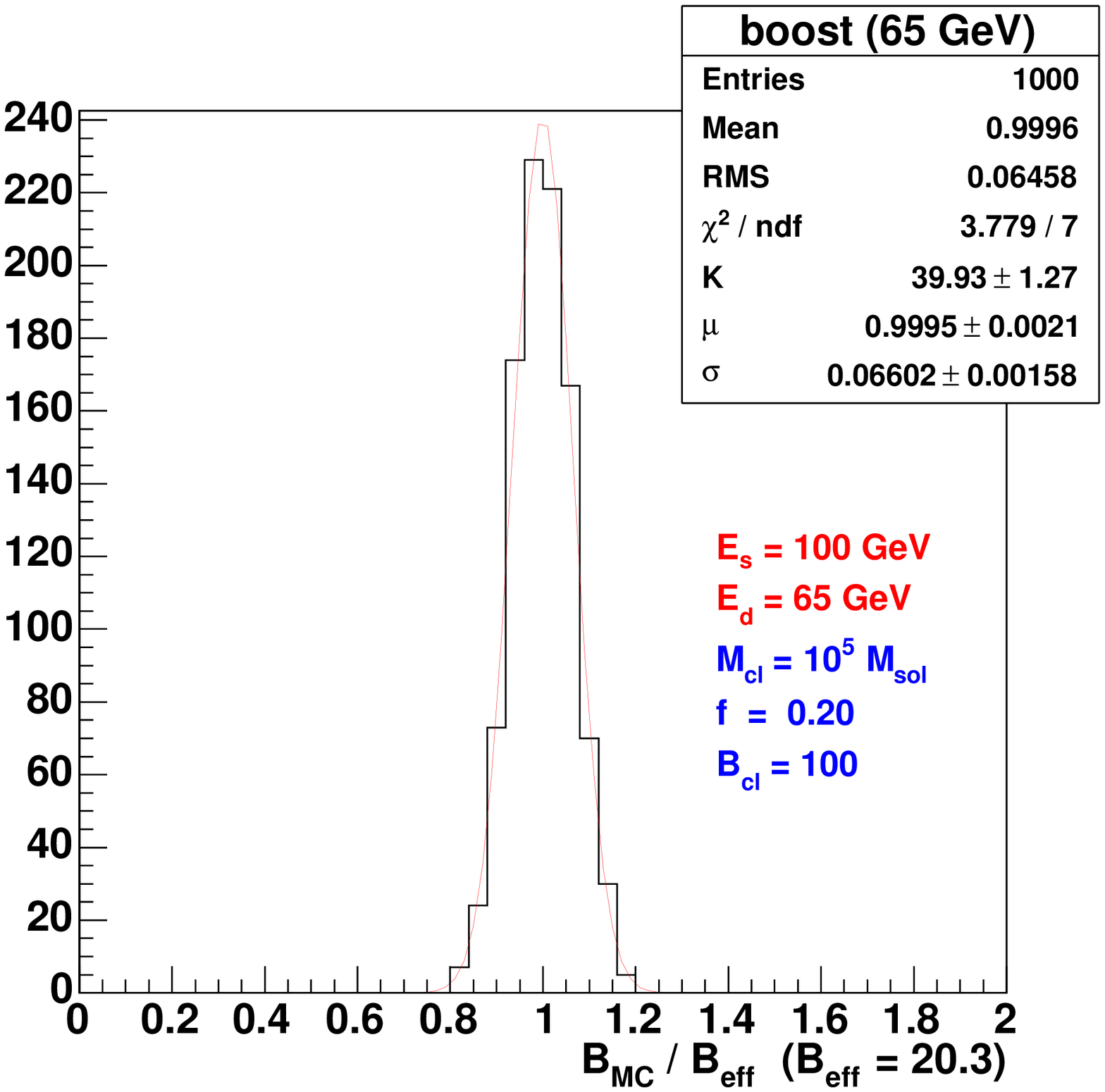}}
  \centerline{\hfill\includegraphics[width=0.5\linewidth]{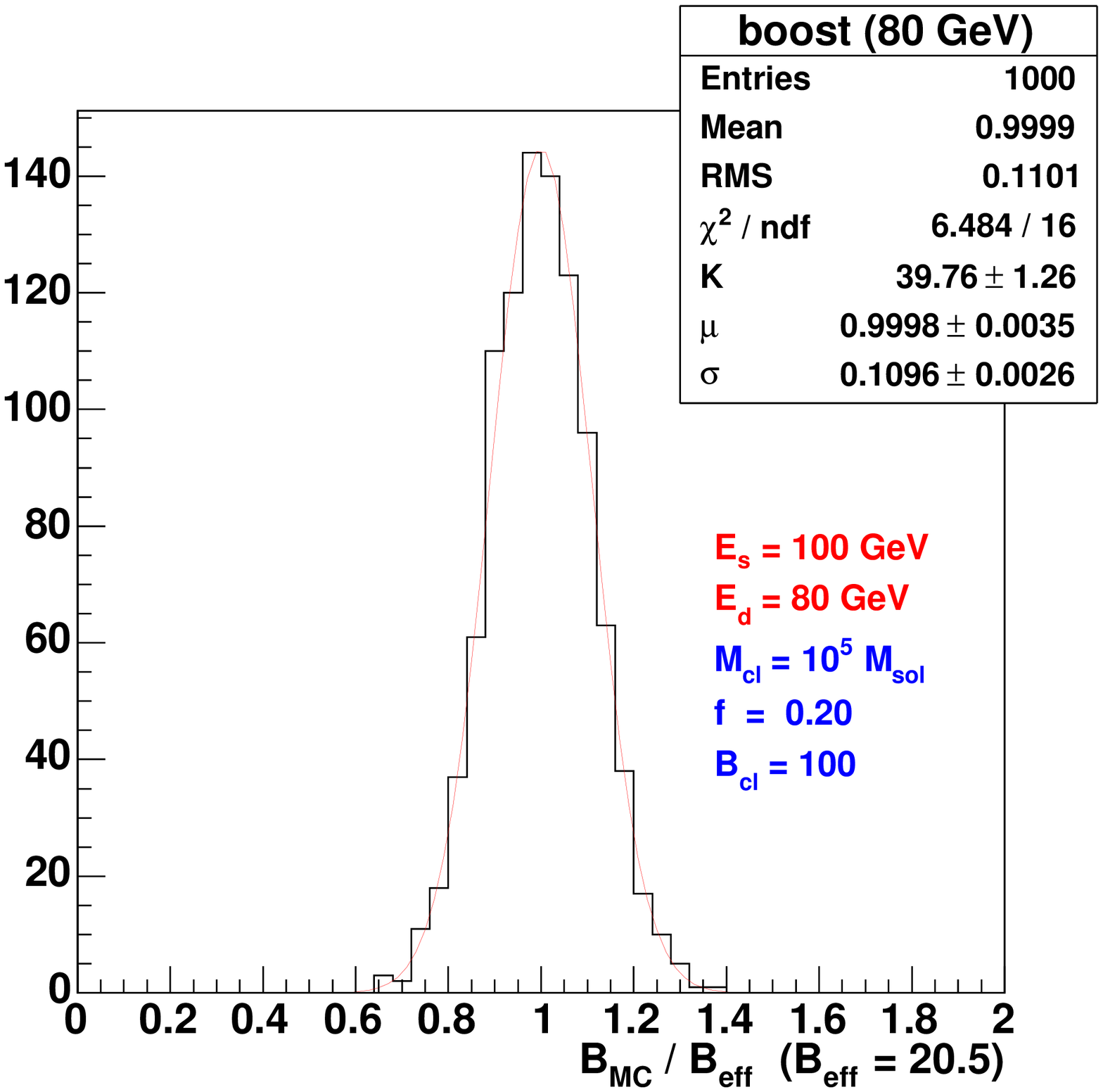}\hfill
    \hfill\includegraphics[width=0.5\linewidth]{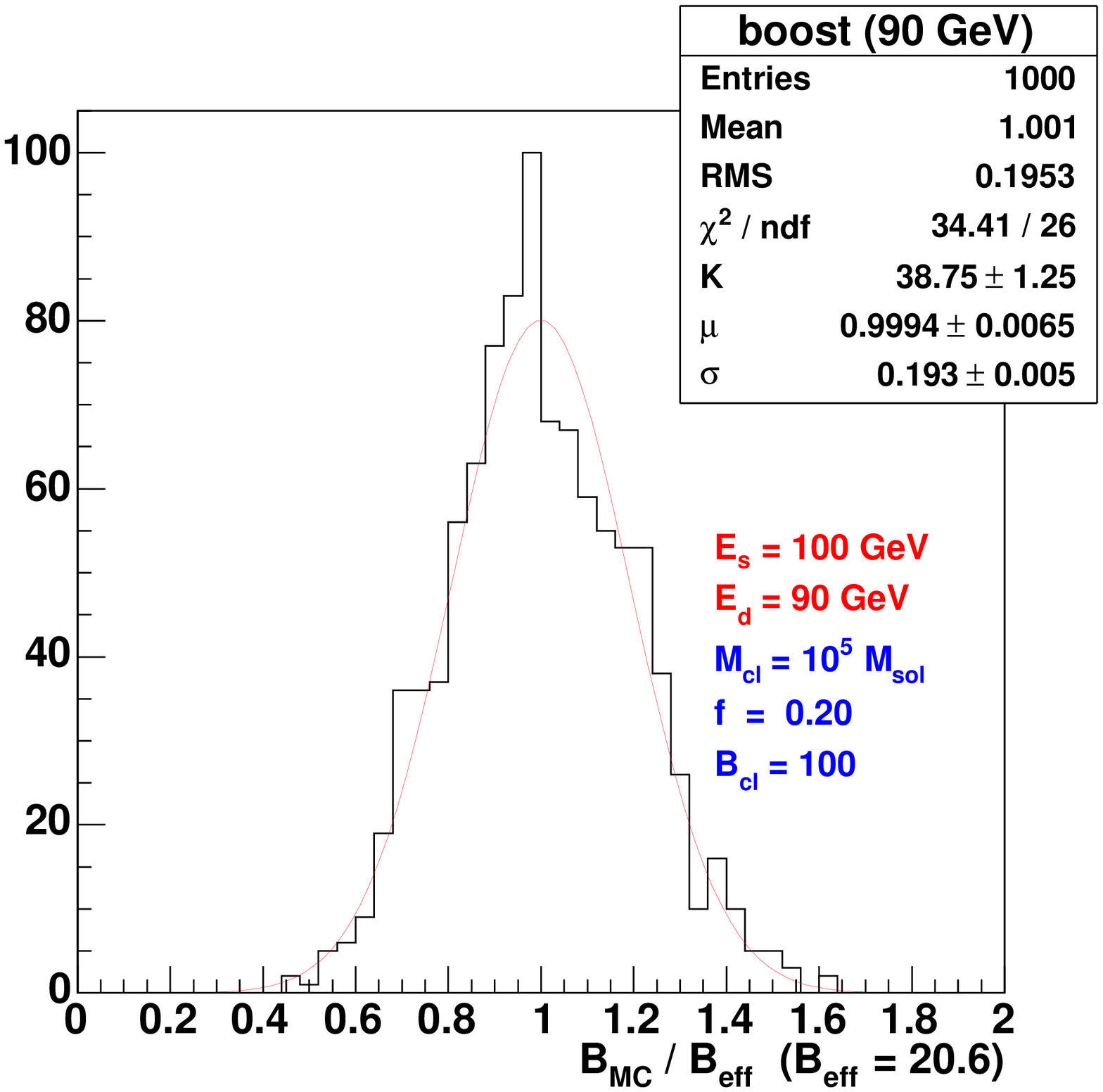}}
  \caption{One thousand different realizations of the distribution of DM substructures
    inside the galactic halo has been generated by Monte-Carlo simulation.
    The injected energy is $E_{S} = 100$ GeV. A NFW profile has been assumed
    with typical scale 25 kpc. A mass fraction $f = 0.2$ is in the form of clumps
    with mass $10^{5}$ M$_{\odot}$.
    Each histogram corresponds to a specific positron energy $E$ at the Earth.
    The number of realizations -- each involving $271,488$ clumps -- is plotted
    as a function of the reduced boost $\eta = {B}/{B_\text{eff}}$.
    Up to an overall factor of a thousand -- that corresponds to the number
    of Monte-Carlo realizations -- each panel features a numerical estimate
    of the probability density $\mathcal{P} \! \left( \eta \right)$.}
  \label{fig:proba_boost_julien}
\end{figure}

When the average number $N_{S}$ of clumps that are involved in the signal
is large, the Poisson statistics~(\ref{poisson_1}) becomes the gaussian
distribution
\begin{equation}
P \left( \delta \right) =
{\displaystyle \frac{1}{\sqrt{\displaystyle 2 \pi N_{S}}}} \,
\exp \left( - \, {\delta^{2}}/{2 N_{S}} \right) \;\; ,
\label{poisson_gaussian_2}
\end{equation}
where $\delta \equiv n - N_{S}$ denotes the departure of the number $n$
of substructures inside the sphere $\mathcal{D}_{S}$ from its average
value $N_{S}$. The associated variance is $\sigma_{n} = \sqrt{N_{S}}$.
We therefore anticipate that the flux $\phi_{r}$ will also be randomly
distributed according to a gaussian law with mean value
$\langle \phi_{r} \rangle$ and variance $\sigma_{r}$.

In order to determine the distribution of probability
$\mathcal{P} \! \left( \phi_{r} \right)$ that drives the random flux
$\phi_{r}$ -- generated by the entire constellation of the clumps lying
inside the reservoir $\mathcal{D}_{H}$ -- we should compute the product
of convolution of the $N_{H}$ distributions of probability
$\mathcal{P} \! \left( \varphi \right)$ associated each to the flux
$\varphi$ of a single substructure -- or alternatively to its reduced flux
$\Phi$ as was discussed in section~\ref{subsec:probability_density}.
Such a task may seem desperate.
%
%
However, in the large $N_{S}$ regime, the central limit theorem
may be fruitfully applied to solve that puzzle. 
This theorem states that the above-mentioned inextricable product of
convolution boils down into a gaussian distribution with
mean value 
$\langle \phi_{r} \rangle \equiv N_{H} \langle \varphi \rangle$
and variance
$\sigma_{r}^{2} \equiv N_{H}
\left\{ \langle \varphi^{2} \rangle \, - \, \langle \varphi \rangle^{2}
\right\}$.
We recognize the expressions~(\ref{average}) and (\ref{variance}) which
have been established and numerically computed in
section~\ref{subsec:random_flux}. Therefore, the probability to obtain
a flux $\phi_{r}$ at the Earth may be expressed as
\begin{equation}
  \mathcal{P} \!
  \left\{ \phi_{r} = \sum_i \, \varphi_{i} \right\}
  = \frac{1}{\sqrt{\displaystyle 2 \pi \sigma_{r}^{2}}} \,
  \exp \left\{ 
    - \frac{\left( \phi_{r} - \langle \phi_{r} \rangle \right)^{2}}
    {2 \sigma_{r}^{2}} \right\} \;\; .
  %
  %
  \label{distribution_proba_phi_r}
\end{equation}
The probability that the total positron flux $\phi$ at the Earth
is enhanced by a factor of $B$ with respect to a completely smooth
DM distribution is readily obtained as
\begin{equation}
  \mathcal{P} \! \left\{ B \equiv \phi / \phi_{s} \right\} =
  \frac{1}{\sqrt{2 \pi \sigma_{B}^{2}}} \,
  \exp \left\{ -
      \frac{\left( B - B_\text{eff} \right)^{2}}{2 \sigma_{B}^{2}} \right\}
    \;\; ,
    \label{distribution_proba_B}
\end{equation}
where the variance $\sigma_{B}$ is given by relation~(\ref{variance_sigma_B}).
Finally the reduced boost $\eta \equiv {B}/{B_\text{eff}}$ follows also
the same gaussian law
\begin{equation}
  \mathcal{P} \! \left\{ \eta \equiv B / B_\text{eff} \right\} =
  \frac{1}{\sqrt{2 \pi \sigma_{\eta}^{2}}} \,
  \exp \left\{ - \, 
    \frac{\left( \eta - 1 \right)^{2}}{2 \sigma_{\eta}^{2}} \right\}
  \;\; ,
  \label{distribution_proba_eta}
\end{equation}
with an average value of
$\langle \eta \rangle = 1$ and a variance
$\sigma_{\eta} = {\sigma_{B}}/{B_\text{eff}}$ not too different from
$\sigma_{r}/\langle \phi_{r} \rangle$ as shown in
formula~(\ref{relative_variances_equal}).

%
\begin{table}[h!]
\begin{center}  
{\begin{tabular}{@{}|c|c|c|c|@{}}
\hline
\hline
$E$ & $B_\text{eff}$ & $\sigma_{\eta} = {\sigma_{B}}/{B_\text{eff}}$ & $N_{S}$ \\
\hline
 50 & 20.09 & 0.04338 & 498.0 \\
 65 & 20.32 & 0.06472 & 223.5 \\
 80 & 20.48 & 0.10966 & 78.0  \\
 90 & 20.57 & 0.19680 & 24.2  \\
\hline
\hline
\end{tabular}}
\end{center}
\caption{
For each value of the positron energy $E$ at the Earth that has been
considered in the plots of Fig.~\ref{fig:proba_boost_julien},
we have computed the corresponding effective boost $B_\text{eff}$ as well
as the variance $\sigma_{\eta}$ associated to the reduced boost
$\eta = {B}/{B_\text{eff}}$. That variance has been derived from
relation~(\ref{relative_variances_equal}) and is in excellent agreement
with the rms value of the Monte-Carlo simulation. The average
number $N_{S}$ of substructures inside the sphere $\mathcal{D}_{S}$
is also indicated.
\label{tab:proba_boost_julien}}
\end{table}
%

\begin{figure}
  \centerline{
    \includegraphics[width=0.45\linewidth]{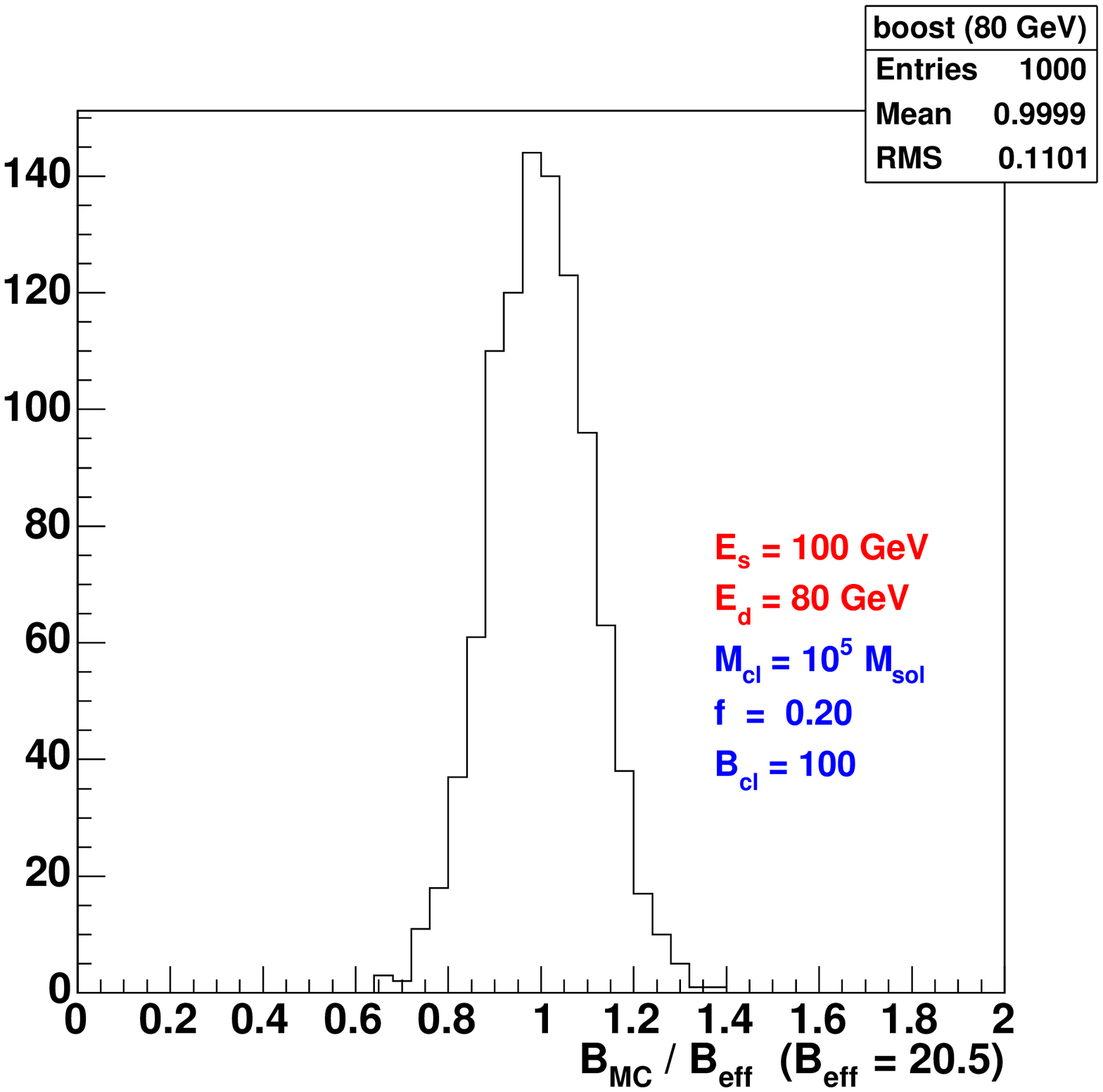}\hfill
    \includegraphics[width=0.45\linewidth]{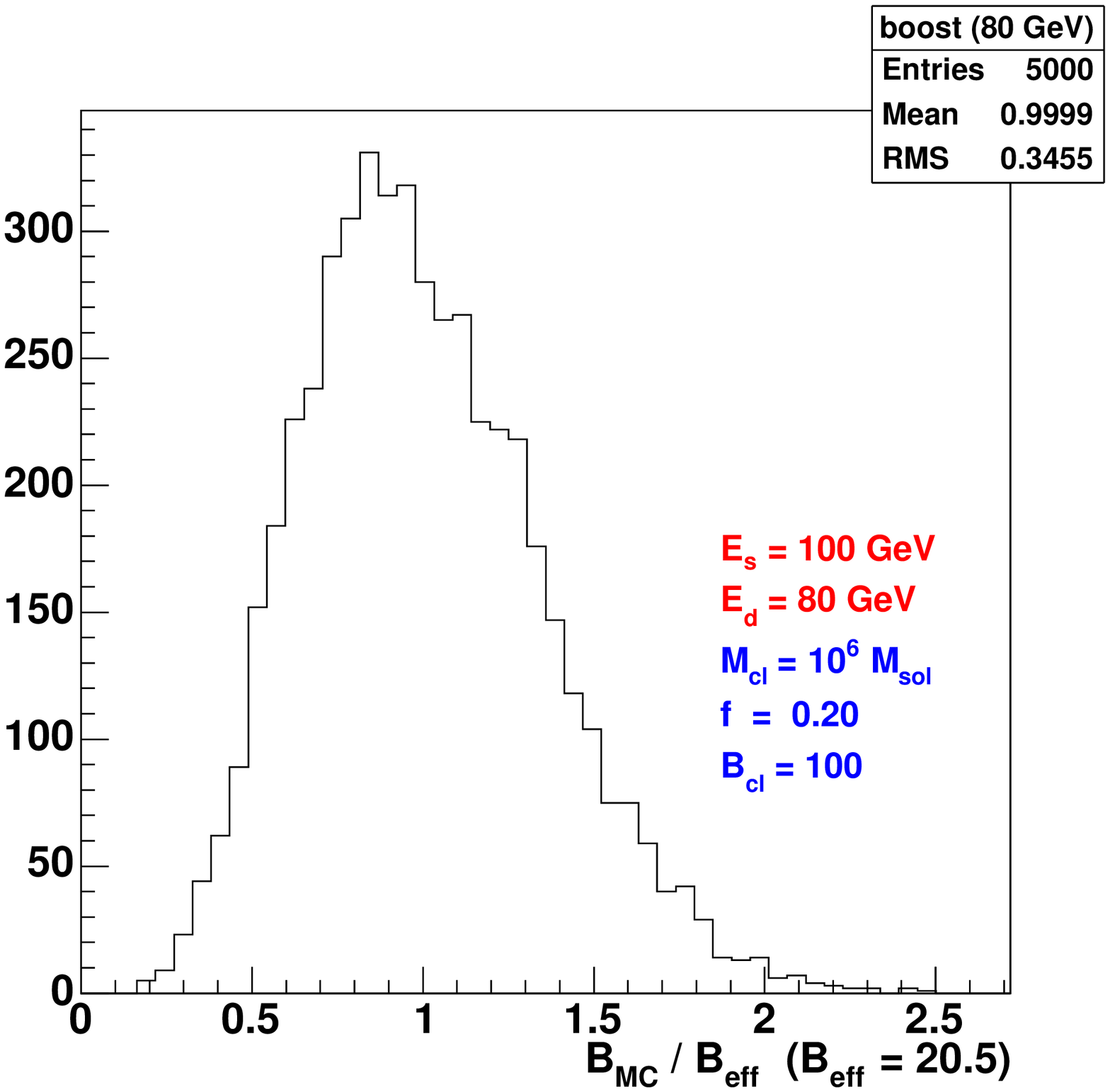}}
  \centerline{
    \includegraphics[width=0.45\linewidth]{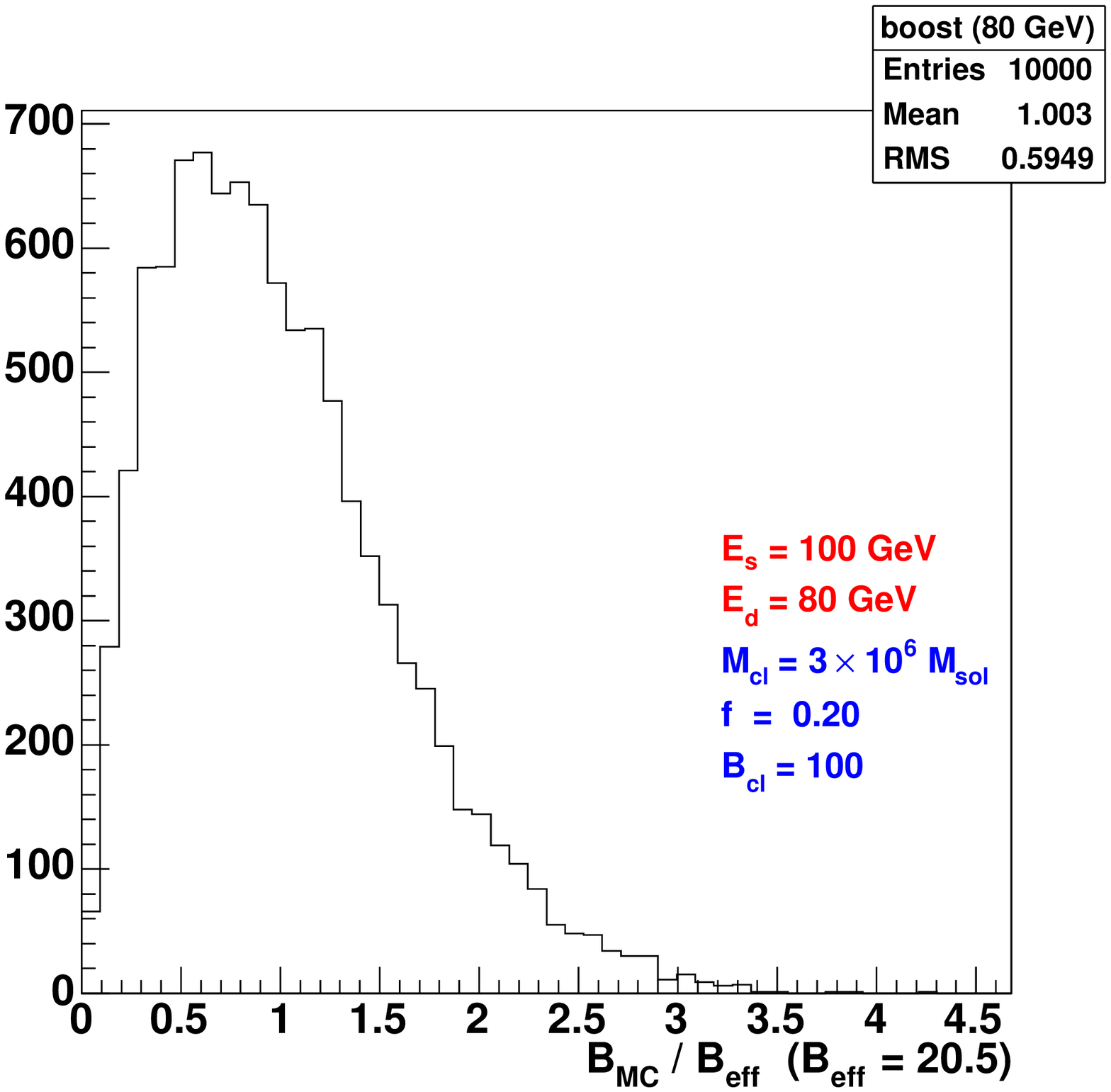}\hfill
    \includegraphics[width=0.45\linewidth]{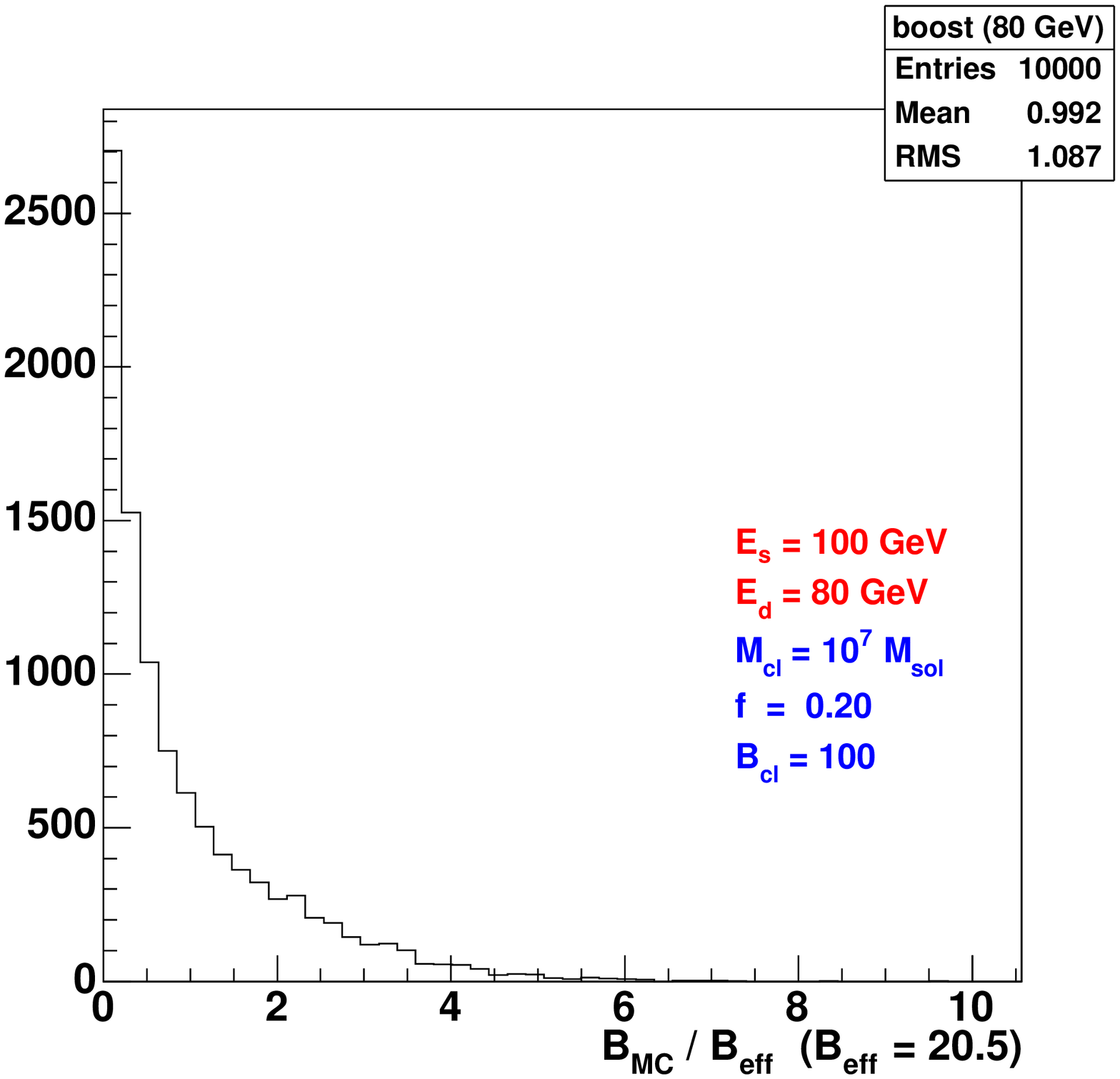}}
  \caption{Distribution of $B_\text{MC}/B_\text{eff}$ for
    an energy $E = 80$ GeV, given the source energy $E_S=100$ GeV, 
    with a mass $M_c$ of
    individual clumps equal to $10^{5}\,M_\odot$,
    $10^{6}\,M_\odot$, $3\times 10^{6}\,M_\odot$ and $10^{7}\,M_\odot$. 
    When $M_c$ is small, the clumps are
    numerous enough for the central limit theorem to apply. 
    The mass fraction in clumps is set to the value $f=0.2$.
    The resulting
    distribution is a gaussian, as described in the text. On the other
    hand, when $M_c$ is large, the probability that several clumps
    contribute to the observed signal is small and the observed
    distribution for $B_\text{MC}/B_\text{eff}$ reflects the one clump
    distribution $P_1(\Phi)$.}
  \label{fig:small_N_limit}
\end{figure}

In order to check our theoretical predictions, we have run a Monte-Carlo
simulation of the distribution of DM substructures in the Milky Way halo.
A thousand different realizations have been generated at random assuming
a NFW DM galactic halo with a fraction $f = 0.2$ in the form of
$10^{5}$ M$_{\odot}$ clumps. In Fig.~\ref{fig:proba_boost_julien}, the
number of realizations is plotted as a function of the reduced boost
$\eta$ for 4 values of the positron energy at the Earth.
These distributions are the Monte-Carlo counterparts of the gaussian
law~(\ref{distribution_proba_eta}) with a mean value of $\eta$ actually
very close to 1. In each panel, the rms value of the histogram is equal
-- within a few percent -- to the variance
$\sigma_{\eta} = {\sigma_{B}}/{B_\text{eff}}$ which we have derived
from expression~(\ref{relative_variances_equal}) and listed in
Tab.~\ref{tab:proba_boost_julien} for comparison with the results of
Fig.~\ref{fig:proba_boost_julien}.
For completeness, each histogram has been independently fitted by the
Gaussian distribution
\begin{equation}
  \mathcal{G} \left( \eta , \mu , \sigma \right) =
  \frac{K}{\sqrt{\displaystyle 2 \pi \sigma^{2}}} \,
  \exp \left\{ - 
      \frac{\left( \eta - \mu \right)^{2}}{2 \sigma^{2}} \right\} \;\; .
\end{equation}
The amplitude $K$, mean value $\mu$ and variance $\sigma$ are displayed
in each panel and the corresponding fitted gaussian is featured by the
red curve. The width of each bin is $\Delta \eta = 0.04$ and since we
have generated $10^{3}$ Monte-Carlo realizations, we should obtain a
value of $K = 0.04 \times 10^{3} = 40$ for the amplitude. This is quite
the case. The mean value $\mu$ of the gaussian is basically equal to 1
whereas its variance $\sigma$ is very close to the Monte-Carlo rms value
and to $\sigma_{\eta}$ -- see Tab.~\ref{tab:proba_boost_julien}.
Because the clumps that are involved in the positron signal at the
Earth are numerous -- the average number $N_{S}$ is still larger
than $\sim$ 20 even at the highest energy $E = 90$ GeV -- the
central limit theorem applies and the gaussian
distribution~(\ref{distribution_proba_eta}) is an excellent description
of the statistical fluctuations of the positron flux.
The question arises now to understand how the
distribution of probability $\mathcal{P} \! \left( \eta \right)$
is modified in the limit where $N_{S}$ becomes smaller than 1.
This is illustrated in Fig.~\ref{fig:small_N_limit}.

\subsection{The small $N_{S}$ regime}
\label{sec:limit_small_N_S}

\begin{figure}
  \centerline{\hfill
    \includegraphics[width=\linewidth]{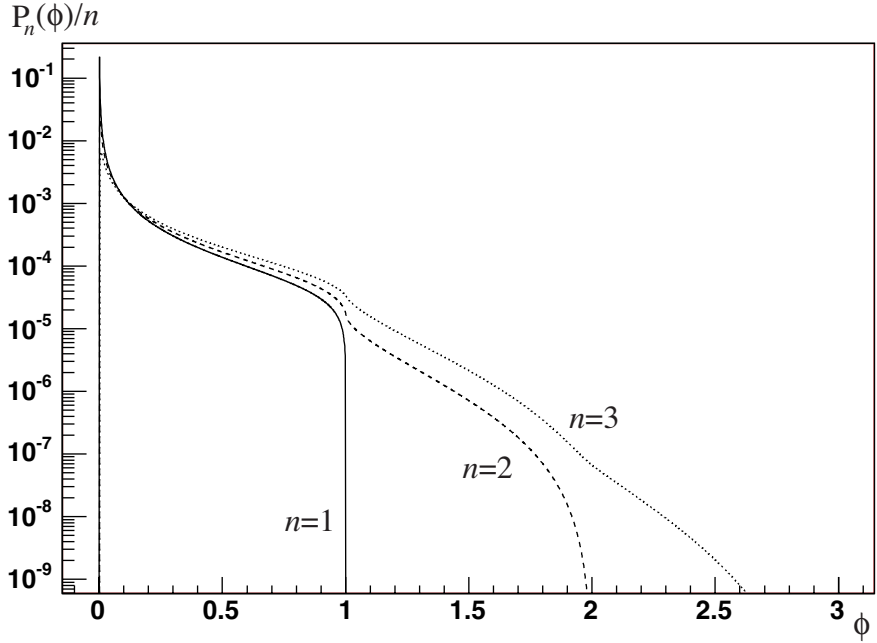}\hfill}
  \caption{Probability distribution $\mathcal{P}_n(\Phi)/n$ for $n=1$, 2 and 3, 
  obtained by consecutive convolutions of $\mathcal{P}_1(\Phi)$.}
  \label{fig:proba_P}
\end{figure}

\begin{figure}
  \centerline{\hfill
    \includegraphics[width=\linewidth]{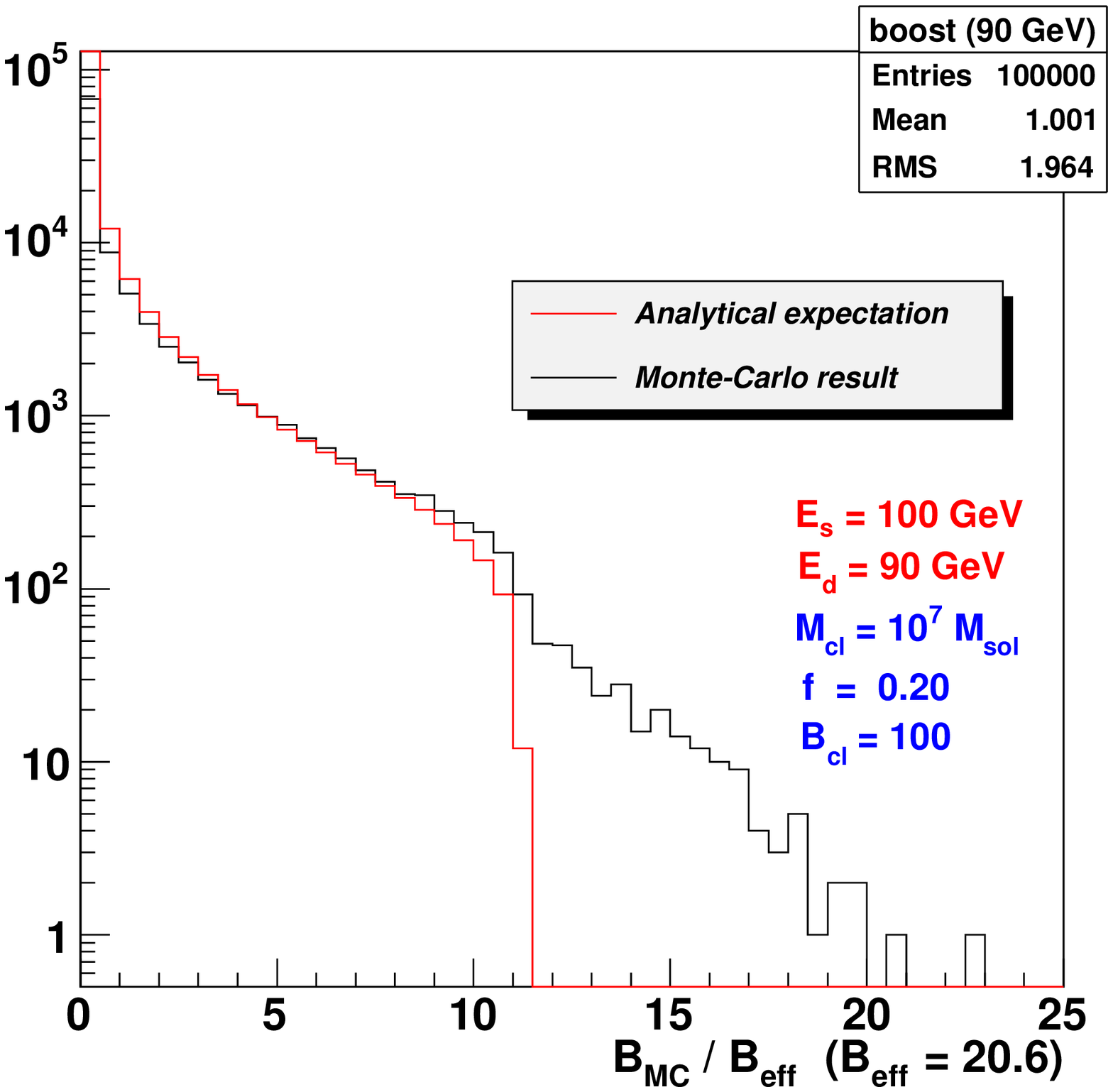}\hfill}
  \caption{Probability distribution of
    $B/B_\text{eff}$ for an assembly of clumps and the corresponding
    distribution for one clump, multiplied by $N$ (see text). 
    The curves are similar in their common range, as is expected
    from an analytical calculation of $\mathcal{P}_N(\Phi)$ 
    (see Fig.~\ref{fig:proba_P}).}
  \label{fig:proba_dist_N}
\end{figure}

When the diffusion range of positrons is
small (for energies close to the emission energy), the individual
probability distribution $\mathcal{P}_1(\Phi) \equiv
\mathcal{P}(\Phi)$ (where $\Phi\equiv \varphi/\varphi_\text{max}$)
is strongly peaked at $\Phi_\text{min}\sim 0$. As a result, the probability
distribution for the total flux $\Phi_\text{tot}$ generated by the $N \equiv N_H$
clumps of domain $\mathcal{D}_H$ can be approximated by
\begin{equation}
  \mathcal{P}_N(\Phi_\text{tot}) = N \mathcal{P}_1(\Phi_\text{tot}) \;\; ,
  \label{low_NS_statistics}
\end{equation}
for $0 < \Phi_\text{tot} < 1$ and in the regime where $\langle \Phi_\text{tot} \rangle$
is vanishingly small.
The proof is straightforward.
The probability $\mathcal{P}_N$ is given by
\begin{equation}
  \mathcal{P}_N(\Phi_\text{tot}) = \int_0^1 \mathcal{P}_1(\Phi)
  \mathcal{P}_{N-1}(\Phi_\text{tot}-\Phi) d\Phi \;\;\;.
\end{equation}
When $\mathcal{P}_{N-1}(\Phi_\text{tot})$ behaves qualitatively
like $\mathcal{P}_1(\Phi)$ and is strongly peaked at a value
close to 0, 
two regions dominate the contribution to the integral when
$\Phi_\text{tot}<1$, namely $\Phi$
close to 0 (where $\mathcal{P}_1(\Phi)$ is large) and $\Phi$ close to
$\Phi_\text{tot}$ (where $\mathcal{P}_{N-1}(\Phi_\text{tot}-\Phi)$ is large), so that
\begin{equation}
  \mathcal{P}_N(\Phi_\text{tot}) \approx \mathcal{P}_{N-1}(\Phi_\text{tot}) + \mathcal{P}_1(\Phi_\text{tot})
\end{equation}
which proves the property~(\ref{low_NS_statistics}).
This is illustrated in Fig.~\ref{fig:proba_P} where the distributions
$\mathcal{P}_1$, $\mathcal{P}_2/2$ and $\mathcal{P}_3/3$ are featured
as a function of the total positron flux expressed in units of the
maximal value $\varphi_\text{max}$ which a single clump can generate
at the Earth. The self convolution of $P_1(\Phi)$ has been carried out
numerically to yield $\mathcal{P}_2$ and $\mathcal{P}_3$, assuming 
relation~(\ref{analytic_mathcal_P_Phi}). The size of the
statistical domain $\mathcal{D}_H$ has been fixed by setting a low flux
cut-off of $\Phi_{min}=0.001$. As is clear in Fig.~\ref{fig:proba_P}, the
three distributions are basically identical in the range where $\Phi$
is smaller than 1. This is so because the probability densities are
peaked at $\Phi = 0$. Should we have chosen a smaller domain $\mathcal{D}_H$
and hence a larger value for the cut-off $\Phi_{min}$, the distributions
would have been less saturated by their low-flux behaviour and
relation~(\ref{low_NS_statistics}) would not have applied.
In Fig.~\ref{fig:proba_dist_N}, $10^5$ realizations of a clumpy DM halo
have been simulated with a substructure mass of $10^{7}$ M$_{\odot}$.
On the horizontal axis, the histogram features the boost 
ratio $\eta \equiv {B}/{B_\text{eff}}$
which is proportional to $\Phi_\text{tot}$. The resemblance with the
analytical distributions of Fig.~\ref{fig:proba_P} is striking. The red
curve which is superimposed on the Monte Carlo results of
Fig.~\ref{fig:proba_dist_N} corresponds to the product
\begin{equation}
  N_{H} \, \mathcal{P}_{1}(\varphi) \equiv
  \frac{f}{M_{c}} \,
  {\displaystyle \int_{\mathcal{D}_{\varphi}}} \, \rho_{s}(\vec{x}) \,
  d^{3} \vec{x} \;\; ,
\end{equation}
with the same values of $f$ and $M_{c}$ as in the simulation. On a
large portion of the range extending from $\sim 0$ up to
$B \sim 11 \, B_\text{eff}$  -- therefore for a total flux smaller than
$\varphi_\text{max}$ -- relation~(\ref{low_NS_statistics}) is a quite
good approximation. This regime corresponds to the situation where
a single clump happens to contribute significantly more than the others
and is the framework of the \cite{cumberbatch06} work.
Most of the realizations of Fig.~\ref{fig:proba_dist_N} correspond to
small values of the flux ratio $\Phi_\text{tot}$.
The number of clumps effectively implied in the signal
is on average very small ($N_{S} \ll 1$).

\section{A practical example: HEAT excess}

The putative positron excess reported by the HEAT experiment around
$\sim 10$ GeV is suggestive of an exotic mechanism such as for
example the annihilation of wimps potentially concealed
inside the halo of the Milky Way.

The present study basically sketches a well defined frame that can be used 
in order to make predictions with respect to available DM candidates and 
experimental data. 
HEAT measurements \citep{heat1,heat2} 
have been widely exploited in connection to annihilating DM, and 
we can easily verify how our results translate into phenomenology. For that 
purpose, we have chosen an illustration based on a DM candidate that was 
proposed in the so-called \emph{warped GUT} theoretical 
scheme \citep{agashe_servant_04}. Within such an 
extra-dimensional modelling, the conservation of a discrete symmetry 
called $Z_3$-symmetry (which is related to the stabilization of the proton, 
like the $R$-parity in supersymmetry), allows the survival of the
lightest $Z_3$-charged particle, which has the properties of a right-handed neutrino
called LZP hereafter. 
This particle is actually a Dirac fermion, and given 
that no matter/anti-matter asymmetry is involved in that case, the 
annihilation rate per volume unit of this species is provided by 
equation~\ref{annihilation_rate} with $\delta=1/4$. We thus have 
considered a fiducial model in which we fixed the LZP mass to 50 GeV, and the 
extra-dimension mass parameter to $m_{KK} = 6$ TeV. The cross-section formulae 
can be found in \citet{agashe_servant_04}, and are mainly 
defined by the isospin content of any final state. It is noticeable that about 
10\% of the annihilation product is carried out equally by the three 
charged lepton/anti-lepton pairs, which can provide a relevant contribution to 
a sharp component close to the wimp mass in the injected positron spectrum. 
We then used the Pythia \citep{pythia} Monte-Carlo to infer the 
positron spectrum associated with all decay and fragmentation processes.

Besides, in order to calculate the expected positron fraction, we made
a further assumption, which asserts that to each positron generated by 
wimp annihilation and propagated to the Earth, an electron is associated with 
the same spectral information. The positron fraction is consequently given 
by the following expression:
\begin{equation}
  f_{e^+}(E) = \frac{\phi_{e^+}(E) + \phi_{e^+,bg}(E)}
  {\phi_{e^+}(E) + \phi_{e^+,bg}(E) + \phi_{e^-}(E) + \phi_{e^-,bg}(E) }
\end{equation}
where $bg$ indicates non-exotic contributions (secondary for 
positrons, both primary and secondary for electrons). For those components, 
we used the estimates by \citet{sm98}.

\begin{figure}
  \centerline{\hfill\includegraphics[width=0.5\linewidth]{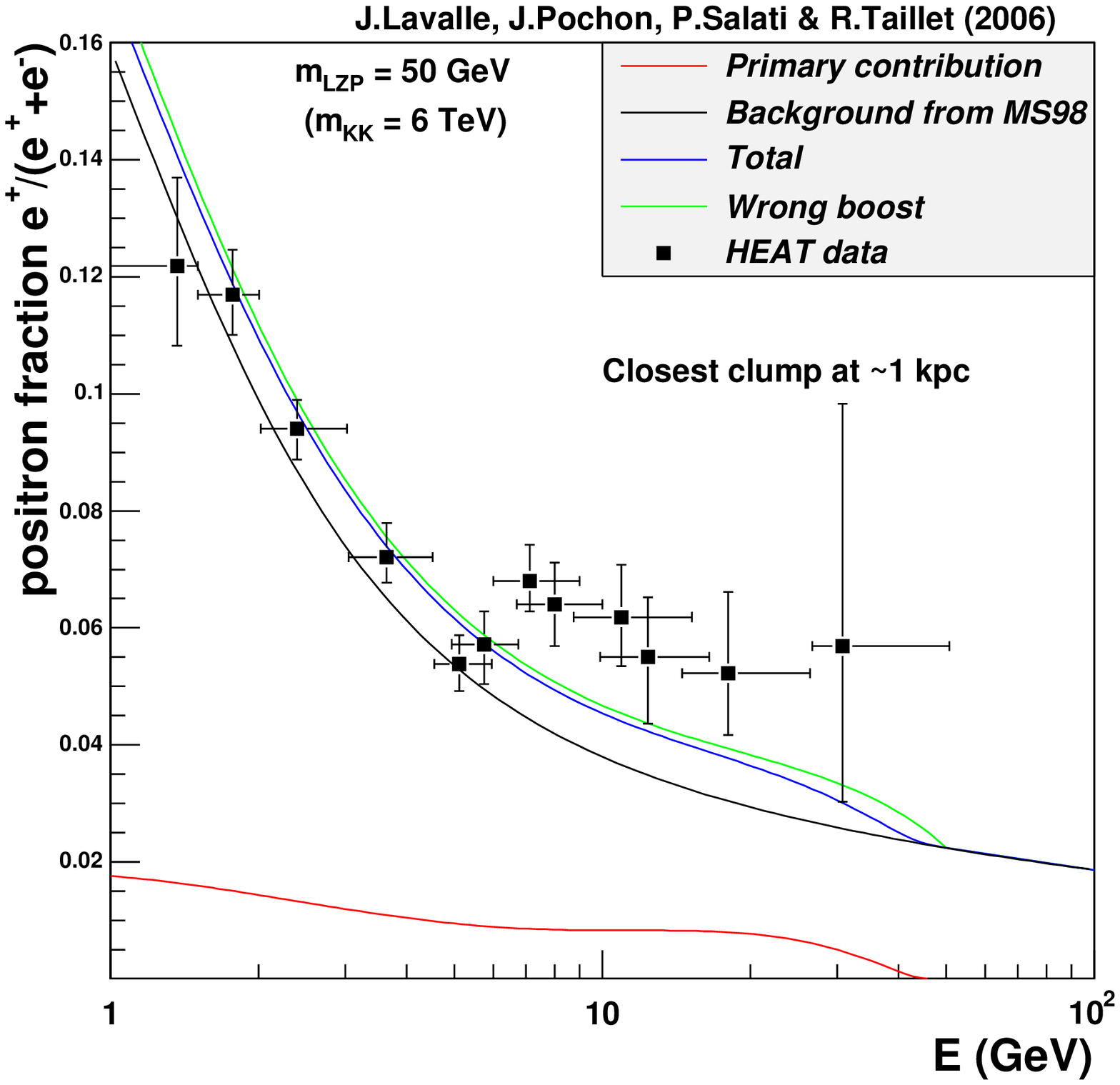}\hfill
    \hfill\includegraphics[width=0.5\linewidth]{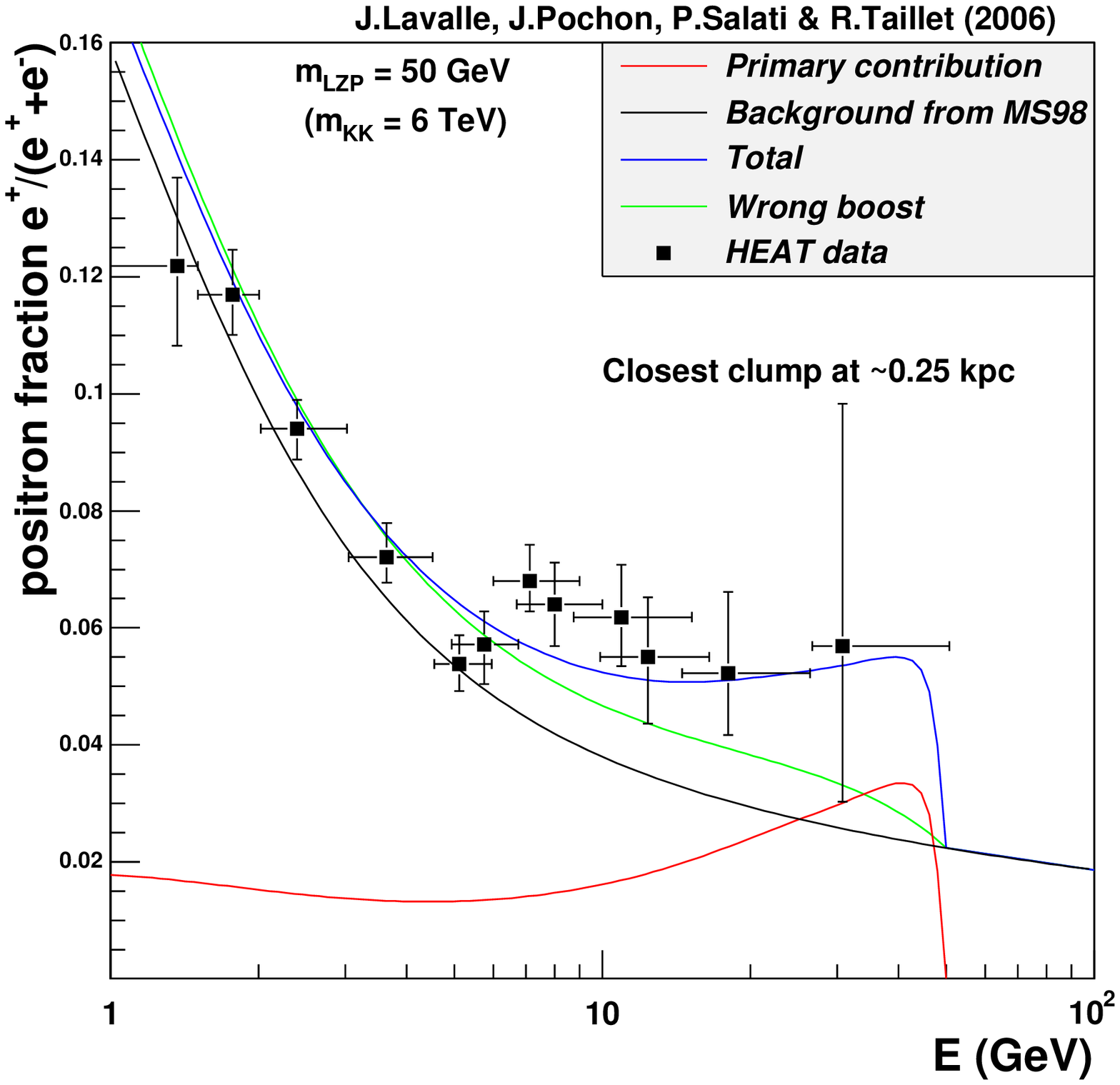}}
  \caption{
    Expected positron fraction in the frame of LZP DM particles, with
    $m_{LZP}=50$ GeV, superimposed on the HEAT data~\citep{heat2}, as
    computed from two random realizations of a clumpy halo with clump
    masses of $10^7 M_{\odot}$. The smooth DM distribution follows an NFW
    profile with a scale radius of 25 kpc, and the mass fraction in clumps
    within 20 kpc has been set to $f=0.2$. An intrinsic clump boost of
    $B_c = 200$ is considered. Left: the closest clump has a distance to
    the Earth of $\sim 1$ kpc. Right: the closest clump has a distance to
    the Earth of $\sim 0.25$ kpc, which is less probable than the
    left-panel case. 
    On both panels, the red and blue curves give respectively the
    primary and the total (primary + background) contributions from DM
    annihilation. A smooth halo would not give rise to any excess with
    respect to the background in that case. For comparison, the green curve
    illustrates the wrong use of the boost, which is simply a shift of the
    spectrum by a factor of $f\times B_c$, 40 in that case.
  }
  \label{fig:heat}
\end{figure}

We show on Fig.~\ref{fig:heat} the results obtained when
considering an NFW profile, and $f=20$\% of the halo mass within a
radius of 20 kpc being clumpy. We will not discuss how the obtained
spectrum is compatible with previous works. Instead, we want to stress
here the differences between the naive account for a global
and \emph{wrong} boost factor set by the product $f\times B_c$, and
the correct treatment of the problem that we have presented in this
paper. To this aim, we used two
particular Monte-Carlo simulations with clump masses of
$10^7 M_{\odot}$ and individual boost factor of $B_c=200$.
This illustrates the situation where a small number of clumps 
contributes, at energies close to $E=m_{LZP}$.
In the left panel of
Fig.~\ref{fig:heat}, the closest clump is found to lie at a distance
of $\sim 1$ kpc to the Earth, that corresponds basically to a regime
in which $B/B_\text{eff}<1$ (\emph{cf.} Fig.~\ref{fig:proba_dist_N}). In
the right panel, the closest clump has a distance to the Earth of $\sim
0.25$ kpc, which is a much less probable configuration, with
$B/B_\text{eff} > 5$. 
The corresponding probability is less than 1\%, as shown by
Fig.~\ref{fig:proba_dist_N}. 
In both panels, the solid blue curves
feature the correct treatment of the boost factor, while the green lines
correspond to the naive shift by a factor of $f\times
B_c$, 40 in this case.
Notice the discrepancy at low energy in both
panels, that is consequent to the energy dependence of the correct
boost factor $B_\text{eff}$.
In the right panel, the Monte Carlo result is five times larger than
the naively boosted flux close to $M_{LZP}=50\,GeV$, 
as a result of the variance affecting this small
$N_s$ configuration.

This indicates how carefully predictions should be made when
computing the flux enhancement due to clumpiness.
We stress that the variance of that boost should also be provided
along with the mean values.
The spectral distorsions could be sizeable 
when compared to the experimental error bars of the HEAT results.
It is of paramount importance to properly take them 
into account when studying the 
discovery potential of the next generation experiments, such
as AMS \citep{ams} or PAMELA \citep{pamela}.
The case of the LZP has been chosen as typical. 
For particles annihilating mostly into charged lepton pairs 
(respectively quark pairs), -- like the lightest Kaluza-Klein 
candidate $B^{(1)}$ of universal extra-dimension theories 
(neutralinos in mSUGRA) -- the effect would be stronger 
(a bit weaker).

Recently, the presence of local dark matter substructure has been
invoked with the specific assumption \citep{cumberbatch06}
that a single neutralino clump would generate alone the observed
distortion in the positron spectrum should it be very close to the
Earth -- at a distance of $\sim$ 0.1 pc. The contribution of the
other and more remote protohalos was assessed to be negligible.

We will not discuss here whether the distortion which that single
substructure generates really matches or not the data nor will we
be interested in the specific nature of the wimp at stake.
The real question which we would like to address is to determine the
probability for a nearby clump alone to shine more strongly than the
rest of the other protohalos. This situation could indeed arise
as suggested by Fig.~\ref{fig:heat}, but its probability
is vanishingly small in the \citet{cumberbatch06} configuration.
The authors of that a priori appealing proposal have assumed
that half of the Milky Way dark matter halo was made of $\sim 10^{15}$
Earth mass clumps as suggested by recent numerical simulations
\citep{Diemand:2005vz}.
That constellation of neutralino substructures is randomly distributed
and contributes on average a positron flux $\langle \phi_{r} \rangle$
at the Earth. The nearby protohalo yields in addition a signal
$\phi_{r} - \langle \phi_{r} \rangle$ that, according to 
\citet{cumberbatch06}, overcomes the contribution
$\langle \phi_{r} \rangle$ from the other clumps. The explanation of
the HEAT excess in terms of that providential protohalo relies therefore
on the assumption that the total positron flux $\phi_{r}$ is larger
than twice the average value $\langle \phi_{r} \rangle$.
The probability for such a configuration may be expressed as
\begin{equation}
P \left\{ \phi_{r} \geq 2 \, \langle \phi_{r} \rangle \right\} =
{\displaystyle \int_{2 \, \langle \phi_{r} \rangle}^{+ \infty}} \,
\mathcal{P} \! \left( \phi_{r} \right) \, d\phi_{r} \;\; .
\end{equation}
The distribution of probability is given by the Maxwell
law~(\ref{distribution_proba_phi_r}) insofar as the central limit 
theorem can be applied in that case. The previous relation
translates into
\begin{equation}
\ln P \left\{ \phi_{r} \geq 2 \, \langle \phi_{r} \rangle \right\}
= - a \, - \, \ln 2 \, - \, \frac{1}{2} \ln \left( \pi a \right)
\;\; ,
\end{equation}
where the parameter $a$ stands for the ratio
$\langle \phi_{r} \rangle^{2} / 2 \sigma_{r}^{2}$. The relevant
statistical quantity which we need to derive is the variance $\sigma_{r}$.

We may readily apply the tools which we have constructed up to the
condition that we are now dealing with a neutralino that
produces a continuous positron spectrum instead of a line. The
positron propagator
$G_{e^+} \left( \vec{x}_{\odot} , E \leftarrow \vec{x} , E_{S} \right)$
needs to be replaced by the convolution
\begin{equation}
  G' \left( \vec{x} \right) \equiv
  G'_{e^+} \left( \vec{x}_{\odot} , E \leftarrow \vec{x} \right) =
  \int_{E}^{m_{\chi}} \,
  G_{e^+} \left( \vec{x}_{\odot} , E \leftarrow \vec{x} , E_S \right)
  \; \left. \frac{dN_{e^+}}{dE_{e^+}} \right|_{E_S}
  \, dE_S \;\; ,
\end{equation}
where ${dN_{e^+}}/{dE_{e^+}}$ denotes the positron spectrum at the
source. Without loss of generality, we have focused our discussion on
a neutralino with mass $m_{\chi} = 100$ GeV and a positron energy at the
Earth of $E = 10$ GeV. 
Neutralinos has been assumed to annihilate into $b \bar{b}$ pairs.
According to the hard-sphere approximation of
section~\ref{subsec:hard_sphere_discussion}, the positrons that are
produced at the energy $E_{S}$ originate from the volume
$V_{S} = ( \sqrt{2 \pi} \, \lambda_{\rm D} )^{3}$
surrounding the Earth. In the case which we consider here, the positrons
that are detected at the energy $E$ at the Earth have been produced at an
energy $E_{S}$ that spans the entire range from $E$ up to the mass
$m_{\chi}$. The volume from which the signal originates on average is
the convolution
\begin{equation}
V'_{S} = {\displaystyle \int_{E}^{m_{\chi}}} \,
V_{S} \left( E , E_{S} \right) \;
\left. {\displaystyle \frac{dN_{e^+}}{dE_{e^+}}} \right|_{E_{S}}
\, dE_{S} \;\; .
\end{equation}
The number of protohalos that contribute to the positron signal at
10 GeV is inferred to be 
\begin{equation}
  N_{S} \simeq \frac{V'_{S} f {\rho_{s}(\odot)}}{M_{c}} \;\; ,
\end{equation}
and the relative variance ${\sigma_{r}}/{\langle \phi_{r} \rangle}$ may
be crudely approximated by ${1}/{\sqrt{N_{S}}}$. With a fraction of clumps
of $f = 0.5$ and a substructure mass of $M_{c} = 10^{-5}$ M$_{\odot}$, we
find a number of protohalos of $N_{S} = 2 \times 10^{13}$ and a relative variance
of $\sigma_{r}/\langle \phi_{r} \rangle \sim 2.22 \times 10^{-7}$.
%
%
Because of that very large number of clumps, the use of the
central limit theorem is plainly justified. The relative variance
of the positron signal is vanishingly small and we therefore anticipate
that the probability
$P \left\{ \phi_{r} \geq 2 \, \langle \phi_{r} \rangle \right\}$
is completely negligible.

The correct calculation of the variance makes use of
relation~(\ref{sigma_r_to_phi_r_F1}) where the integrals $\mathcal{I}_{1}$
and $\mathcal{J}_{1}$ are now computed with the convoluted positron
propagator $G' \left( \vec{x} \right)$. We have derived a value of
$\sigma_{r}/\langle \phi_{r} \rangle \sim 4.13 \times 10^{-7}$
in good agreement with the hard-sphere approximation. We therefore
reach the conclusion that
%
%
\begin{equation}
  \log_{10} P \left\{ \phi_{r} \geq 2 \, \langle \phi_{r} \rangle \right\}
  = - 1.27 \times 10^{12} \;\; .
\end{equation}
With such an exceedingly small value of the probability, the
configuration in which a single clump overcomes the signal from the
other $2 \times 10^{13}$ substructures is completely unlikely and
the hypothesis pursued in \citet{cumberbatch06} should be
abandoned.
This example illustrates how the tools which we have presented in this
article may be fruitfully applied in order to derive quantitative results
and not just mere qualitative arguments.


\section{Discussion and conclusion}
 
\paragraph{Summary  :} The enhancement of indirect signals
coming from dark matter annihilation in a clumpy halo is usually
described by a ``boost factor''. We have shown that this quantity
should be considered as a random variable and we have investigated its
statistical properties in the following situation:
(i) clumps are distributed like the smooth component; 
(ii) all the clumps are identical.
We showed that the boost factor may strongly depend upon the specific
realization of the clumpy halo we are living in.

\paragraph{Spatial distribution of clumps:} A more realistic model would include the mass profile of
these clumps, their own density shape and geometry, as well as a
number distribution inspired by hierarchical structure formation
studies (which is often found to be close to the smooth distribution,
at least at large galactic radii~\citep{berezinsky03}). Taking another
number distribution would essentially modify the shape of the
effective boost factor (the integral $\mathcal{I}_1$ of
eq.~\ref{B_eff_19}), whereas individual clump properties
would affect mostly our estimates of its variance at short scales. For
example, the clump number distribution is very likely to be cut off
inside the galactic bulge because of strong tidal interactions with
stars~\citep{berezinsky05}. 
However, we do not expect that the results presented
here would be strongly affected by these effects.
For instance, we have shown that the steep energy
dependence of the mean boost factor, in the case of a positron line,
was mostly due to contributions from our very local environment, due
to the short scale of positron propagation. A cut-off in the clump
distribution for galactic radii less than $\sim 3$ kpc (corresponding
to a minimum of 5 kpc from the Earth) would then significantly
diminish its low energy contribution, and would thus increase the
relative variation of the boost factor with energy. 

\paragraph{Cumberbatch \&Silk:} It has been proposed recently
\citep{cumberbatch06} that the positron excess observed by HEAT
could be due to the presence of a single clump located near 
the Earth. However, the situation in which the signal due to one clump
dominates over the background due to all the others is very unlikely.
The probability that a clump lies in the close proximity of Earth is
sizeable only if the density of clumps is high, which in turn implies
that many of them contribute to the measured flux. 
We showed that the quantitative study of this situation leads to
unreasonably small probabilities. 

\begin{acknowledgements}
  We thank the PNC (french Programme National de Cosmologie) for financial
  support to this work. J.L. is grateful to Alain Falvard and LPTA-Montpellier,
  and to Paschal Coyle and the ANTARES group at CPPM for having supported his 
  contribution. J.P. thanks Jacques Colas and LAPP for extra financial support.
  Part of this work was discussed during the GdR-susy-PCHE meetings. 
  We also thank the anonymous referee for his thorough work on
  the first version of this paper and his very helpful comments.
\end{acknowledgements}


\bibliographystyle{aa}
\bibliography{papier_ps}

\end{document}